\documentclass[authoryear,5p,times]{elsarticle}

\usepackage{amssymb}

\usepackage{rotating}

\journal{Icarus}

\newcommand{\degr}{\ensuremath{^\circ}}
\newcommand{\arcsec}{\mbox{\ensuremath{^{\prime\prime}}}}

\usepackage{hyperref}
\hypersetup{
    bookmarks=true,         
    unicode=false,          
    pdftoolbar=true,        
    pdfmenubar=true,        
    pdffitwindow=false,     
    pdftitle={Physical Properties of (2) Pallas},
    pdfauthor={Benoit Carry}, 
    pdfsubject={Planetary Science}, 
    pdfkeywords={},         
    pdfnewwindow=true,      
    colorlinks=true,        
    linkcolor=gray,         
    citecolor=blue,         
    filecolor=gray,         
    urlcolor=gray           
}

\usepackage{ulem}

\newcommand{\add}[1]{#1}

\usepackage{graphicx}
\graphicspath{{figures/}}

\begin{document}

\begin{frontmatter}



\title{Physical Properties of (2) Pallas\tnoteref{1}}
\tnotetext[1]{Based on observations
  collected at the European Southern Observatory (ESO), Paranal, Chile -
   \href{http://archive.eso.org/wdb/wdb/eso/sched_rep_arc/query?progid=074.C-0502}{074.C-0502}
   \& \href{http://archive.eso.org/wdb/wdb/eso/sched_rep_arc/query?progid=075.C-0329}{075.C-0329}
  and at the W. M. Keck Observatory, which is operated as a scientific
  partnership among the California Institute of Technology, the
  University of California and the National Aeronautics and Space
  Administration. The Observatory
  was made possible by the generous financial support of the W. M. Keck
  Foundation.}

\author[eso,lesia]{Beno\^{i}t Carry}
\ead{benoit.carry@obspm.fr}
\author[eso]{Christophe Dumas}
\author[mikko]{Mikko Kaasalainen}
\author[imcce]{J\'er\^{o}me Berthier}
\author[swri]{William J. Merline}
\author[lesia]{St\'ephane Erard}
\author[keck]{Al Conrad}
\author[aflr]{Jack D. Drummond}
\author[imcce]{Daniel Hestroffer}
\author[lesia]{Marcello Fulchignoni}
\author[onera]{Thierry Fusco}

\address[eso]{ESO, Alonso de C\'{o}rdova 3107, Vitacura, Casilla 19001, Santiago de Chile, Chile}
\address[lesia]{LESIA, Observatoire de Paris, CNRS, 5 place Jules Janssen, 92190 Meudon Cedex, France}
\address[mikko]{P.O. Box 68 (Gustaf H\"{a}llstr\"{o}min katu 2b), FI-00014 University of Helsinki, Finland}
\address[imcce]{IMCCE, Observatoire de Paris, CNRS, 77 av. Denfert Rochereau, 75014 Paris, France}
\address[swri]{Southwest Research Institute, 1050 Walnut St. \# 300, Boulder, CO  80302, U.S.A.}
\address[keck]{W. M. Keck Observatory, Hawaii, U.S.A.}
\address[aflr]{Starfire Optical Range, Directed Energy Directorate,
  Air Force Research Laboratory, Kirtland AFB, New Mexico 87117-577, U.S.A.}
\address[onera]{ONERA, BP 72, 92322 Ch\^{a}tillon Cedex, France}

\begin{abstract}
\setlength\parindent{16pt}
  \indent Ground-based high angular-resolution images of asteroid (2)
  Pallas at near-infrared wavelengths have been used
  to determine
  its physical properties (shape, dimensions, spatial
  orientation and albedo distribution).\\
  \indent We acquired and analyzed adaptive-optics
  (AO)
  \textsl{J/H/K}-band observations from Keck II and the Very
  Large Telescope taken during four Pallas oppositions 
  between 2003 and 2007, with
  spatial resolution
  spanning 32--88 km
  (image scales 13--20 km/pix).
  We improve our determination of the size, shape, and pole
  by a novel method that combines our AO data 
  with 51 visual light-curves
  spanning 34  years of observations
  as well as archived occultation data.\\
  \indent The shape model of Pallas derived here
  reproduces well both the projected shape of
  Pallas on the sky (average deviation
  of edge profile
  of 0.4 pixel) and light-curve behavior
  (average deviation of 0.019 mag) at all
  the epochs considered.
  We resolved the pole ambiguity and found the spin-vector coordinates
  to be within 5\degr~of [long, lat] = 
  [30\degr, -16\degr] in the Ecliptic J2000.0 reference frame,
  indicating a
  high obliquity of about 84\degr, leading to high seasonal contrast.
  The best triaxial-ellipsoid fit returns ellipsoidal radii of
  $a$=275 km, $b$= 258 km, and $c$= 238 km.
  From the mass of Pallas determined by gravitational perturbation
  on other minor bodies [(1.2 $\pm$ 0.3) $\times$ 10$^{-10}$
      M$_\odot$, Michalak 2000, A\&A, 360],
  we derive a density
  of 3.4 $\pm$ 0.9 g.cm$^{-3}$
  significantly different from
  the density of
  C-type (1) Ceres of 2.2 $\pm$ 0.1 g.cm$^{-3}$ [Carry et al. 2008,
    A\&A, 
    478]. 
  Considering the 
  spectral similarities of Pallas and Ceres
  at visible and near-infrared 
  wavelengths, this may point to fundamental differences in the interior 
  composition
  or structure of
  these two bodies. \\
  \indent We define a planetocentric longitude system for Pallas,
   following IAU guidelines. We also present the first albedo maps of Pallas
  covering $\sim$80\% of the surface in K-band.
  These maps reveal
  features with diameters in the 70$-$180 km range and an albedo
  contrast of about 6\% with respect to the mean surface albedo.
\end{abstract}

\begin{keyword}


ASTEROIDS\sep
ADAPTIVE OPTICS\sep
INFRARED OBSERVATIONS\sep
ASTEROIDS, SURFACES\sep
OCCULTATIONS
\end{keyword}

\end{frontmatter}


\section{Introduction}
  A considerable amount of information regarding the
  primordial planetary processes that occurred during and immediately
  after the accretion of the early planetesimals is still present among
  the population of small solar system bodies
  \citep{2002-AsteroidsIII-1-Bottke}.\\
  %
  %
  \indent Fundamental asteroid properties
  include
  composition (derived from spectroscopic analysis) and
  physical parameters (such as size, shape, mass,
  and spin orientation).
  While compositional investigations
  can provide
  crucial information on the conditions in
  the primordial solar nebula \citep{2007-AREPS-35-Scott}
  and on asteroid thermal
  evolution \citep{1990-Icarus-88-Jones}, the study of
  asteroid physical properties 
  can yield insights on
  asteroid
  cratering history \citep{1999-Icarus-140-Davis},
  internal structure \citep{2002-AsteroidsIII-4.2-Britt}, 
  and volatile fraction \citep{2008-MNRAS-383-Mousis} 
  for example.
  These approaches complement one another --- the density
  derived by observations of physical properties
  strongly constrains the composition
  \citep{2002-AsteroidsIII-2.2-Merline}, which is key
  to evaluation of
  evolution scenarios.\\
  \indent Spacecraft missions to asteroids,
  for example
  NEAR to (253) Mathilde and (433) Eros
  \citep{1999-Icarus-140-Veverka}, and
  Hayabusa to (25143) Itokawa \citep{2006-Science-312-Fujiwara},
  greatly enhanced
  our understanding of asteroids.
  The high cost of space missions, however, 
  precludes exploration of more than a few asteroids, leaving
  most asteroids to be studied
  from Earth-based telescopes. \\
  %
  %
   \indent Although several remote observation techniques can be used to
   determine the
   physical properties of asteroids, 
   our technique relies primarily
   on disk-resolved observations.
   Indeed, knowing accurately the size is crucial for the determination of asteroid
   volume, and hence density.
   If enough chords are observed, 
   occultations provide precise measurement
   of asteroid shape and size \citep{1989-AsteroidsII-Millis}, 
   but at only one rotational phase (per occultation event).
   Moreover, because occultations of bright stars seldom occur, 
   only
   a small fraction of all
   occultations are 
   covered by a significant number of observers.
   Describing an asteroid's 3-D
   size and shape with this method
   thus requires decades.
   Assuming a tri-axial ellipsoidal shape is a common way to
   build upon limited observations of asteroid projected sizes
   \citep{1989-Icarus-78-Drummond}.
    From the inversion of photometric light-curves,
    one can also derive asteroid shapes
    \citep{2002-AsteroidsIII-2.2-Kaasalainen}, with
    sizes then relying
    on albedo considerations.
    On the other hand, disk-resolved observations, either radar or
    high angular-resolution imagery, 
    provide direct measurement of
    an asteroid's
   size and shape when its
   apparent disk can be spatially resolved.\\
   \indent For about a decade now, we have
   had access to
   instrumentation with the angular resolution
   required to spatially resolve
   large main-belt asteroids at
   optical wavelengths.
   This can be done in the visible from space with the Hubble
   Space Telescope (HST) or in near-infrared from large telescopes
   equipped with adaptive optics 
   (AO) such as Keck, the Very Large Telescope (VLT),
   and Gemini.
   Disk-resolved observations allow direct measurement of an asteroid's absolute
   size \citep{1993-Icarus-105-SaintPe2}, and of its shape, if enough
   rotational-phase coverage is obtained
   \citep{2007-Science-316-Taylor,2007-Icarus-191-Conrad}.
   One can also derive the spin-vector
   coordinates from the
   time evolution of limb contours
   \citep{2005-Nature-437-Thomas} or from the
   apparent movement of an albedo feature \citep{2008-AA-478-Carry}.
   Ultimately, albedo maps may provide significant
   constraints
   on surface properties such as mineralogy
   or degree of space
   weathering \citep{1997-Icarus-128-Binzel, 2006-Icarus-182-Li,
     2008-AA-478-Carry}.\\
   \indent Even if images can provide a complete description of
   asteroid properties, their combination with other sources of data
   (like light-curves or occultations) can significantly improve
   asteroid 3-D shape models
   \citep[see the shape model of (22) Kalliope in][for
     instance]{2008-Icarus-196-Descamps}. \\  
  %
  \indent Pallas is a B-type asteroid \citep{2002-Icarus-158-BusII}.
  As such, it is thought to have a composition similar
  to that of the
  Carbonaceous Chondrite (CC) meteorites \citep[see][for a
    review]{1983-Icarus-56-Larson}.
  Spectral analysis of the 3 micron band \citep{1990-Icarus-88-Jones}
  exhibited by Pallas 
  suggests that its surface has a significant anhydrous component
  mixed with 
  hydrated CM-like silicates (CM is a subclass of CC
    meteorites).
  Although Pallas is generally linked to CC/CM material, 
  its composition remains uncertain.
  Indeed, Pallas'
  visible and near-infrared spectrum is almost flat with
  only a slight blue slope,
  with
  the only absorption band clearly detected
  being the 3 micron band.\\
  \indent Compositional/mineralogical studies
  for Pallas are further hampered by a poorly determined density.
  First,
  there is significant uncertainty in the mass,
  as most
  mass estimates do not overlap within the error bars
  \citep[see][for a review]{2002-AsteroidsIII-2.2-Hilton}.
  Second, although the size of Pallas has been
  estimated
  from two
  occultations \citep{1979-AJ-84-Wasserman,1990-AJ-99-Dunham},
  at least three events are required to determine 
  asteroid spin and tri-axial dimensions
  \citep{1989-Icarus-78-Drummond}. \\
  \indent Until recently,
  the only published 
  disk-resolved observations of Pallas were limited to some 
  AO snapshots collected in 1991 by \citet{1993-Icarus-105-SaintPe1}, 
  but the lack of spatial resolution prevented
  conclusions about Pallas'
  size, shape, or spatial orientation.
  Recent observations of Pallas from Lick \citep{2008-Icarus-197-Drummond} and
  Keck Observatories \citep{2009-Icarus-Drummond} lead to new
  estimates for its triaxial ellipsoid dimensions, but there was still a
  relatively large uncertainty on the short axis. These Keck
  observations are included as a subset of the data considered here.
  Also, observations of Pallas were recently obtained using 
  the WFPC2 instrument on HST 
  \citep[see][]{2009-LPI-Schmidt}.

\section{Observations}
  \indent Here we present
  high angular-resolution images of asteroid (2) Pallas,
  acquired at multiple epochs, using AO in the near infrared with
  the Keck II telescope and
  the ESO Very Large Telescope (VLT).\\
  \indent During the 2003, 2006 and 2007 oppositions, 
  we imaged Pallas in Kp-band [central wavelengths and bandwidths
      for all bands are given in Table~\ref{tab-obs-settings}] with a 9.942
  $\pm$ 0.050 milliarcsec per pixel image scale of NIRC2, the second
  generation near-infrared camera (1024$\times$1024 InSb Aladdin-3) and
  the AO system installed at the Nasmyth focus of
  the Keck II telescope \citep{2004-AppOpt-43-vanDam}.
  We acquired five other epochs near the more favorable 2005
  opposition 
  during which
  we imaged Pallas in J-, H-, and Ks-bands,
  with the 13.27 $\pm$ 0.050 milliarcsec per pixel image scale
  of CONICA (1024$\times$1026 InSb Aladdin-3)
  \citep{2003-SPIE-4839-Rousset,2003-SPIE-4841-Lenzen} and the NAOS
  AO system installed at the Nasmyth B focus of UT4/Yepun at the VLT.
  We list in Table~\ref{tab-obs-condition} Pallas' heliocentric
  distance and range to observer,
  phase angle, angular diameter and Sub-Earth-Point
  (SEP, with planetocentric coordinate
  system defined in section~\ref{subsec-mapping-method})
  coordinates for each observation. \\
  \indent Near-infrared broad-band
  filter observations of Pallas were interspersed with
  observations of a
  Point-Spread-Function (PSF) reference star at similar airmass and
  through the same set of filters (Tables~\ref{tab-obs-settings}
  \&~\ref{tab-obs-psf}). This calibration was required to 
  perform \textsl{a posteriori} image restoration (deconvolution) as
  described in \citet{2008-AA-478-Carry}.
  These observations of stars also can be used to measure the
  quality of the AO correction during the observations. We thus
  report in Table~\ref{tab-obs-psf} the Full Width at Half Max (FWHM)
  of each PSF, in milliarcseconds and also in kilometers
  at the distance of Pallas.
  No offset to sky was done,
  but the telescope position was dithered
  after one or a few exposures to place the object
  (science or calibration)
  at three different locations on the
  detector separated by
  $\sim$5\arcsec~from each other. 
  This allows a median sky frame to be created directly from
  the acquired targeted images.
 %
 %

 %
\begin{table*}
\begin{center}
  \textbf{Observation Conditions}\\
  \begin{tabular}{cccccccccccc}
    \hline\hline
    Date & UT & $\Delta$ & $r$ & V & $\alpha$ & $\phi$ & SEP$_\lambda$ & SEP$_\varphi$ & Airmass & PSF \\
         &    & (AU) & (AU) & (mag.) & (\degr) & (\arcsec)  & (\degr) & (\degr)     &  & (Table~\ref{tab-obs-psf})\\
    \hline
    2003 Oct 10 & 12:00 & 2.73 & 1.80 & 8.25 & \textcolor{white}{0}9.4 & 0.39 & 107 & -76 & 1.28 & Oct.10-$\star$1\\
    2003 Oct 12 & 09:13 & 2.73 & 1.80 & 8.24 & \textcolor{white}{0}9.5 & 0.39 & 183 & -75 & 1.40 & Oct.12-$\star$1 \\
    2003 Oct 12 & 11:14 & 2.73 & 1.80 & 8.24 & \textcolor{white}{0}9.5 & 0.39 & \textcolor{white}{0}90 & -75 & 1.25 & Oct.12-$\star$2 \\
    2005 Feb 02 & 06:30 & 2.27 & 1.60 & 8.04 & 21.9 & 0.44 & 265 & +64 & 1.21 & Feb.02-$\star$1 \\
    2005 Feb 02 & 08:05 & 2.26 & 1.60 & 8.04 & 21.9 & 0.49 & 192 & +64 & 1.05 & Feb.02-$\star$2 \\
    2005 Mar 12 & 06:02 & 2.34 & 1.37 & 7.20 & \textcolor{white}{0}6.9 & 0.52 & \textcolor{white}{0}54 & +64 & 1.15 & Mar.12-$\star$ \\
    2005 Mar 13 & 04:42 & 2.34 & 1.37 & 7.18 & \textcolor{white}{0}6.6 & 0.52 & \textcolor{white}{0}90 & +64 & 1.21 & Mar.13-$\star$ \\
    2005 May 08 & 23:30 & 2.47 & 1.77 & 8.39 & 20.1 & 0.40 & 326 & +54 & 1.74 & May.08-$\star$1 \\
    2005 May 09 & 23:18 & 2.47 & 1.78 & 8.41 & 20.3 & 0.40 & 309 & +54 & 1.80 & May.09-$\star$1 \\
    2006 Aug 16 & 06:55 & 3.35 & 2.76 & 9.85 & 15.5 & 0.26 & \textcolor{white}{0}22 & +32 & 1.00 & Aug.16-$\star$1 \\
    2006 Aug 16 & 07:22 & 3.35 & 2.76 & 9.85 & 15.5 & 0.26 & \textcolor{white}{00}1 & +32 & 1.01 & Aug.16-$\star$1 \\
    2006 Aug 16 & 07:45 & 3.35 & 2.76 & 9.85 & 15.5 & 0.26 & 343 & +32 & 1.03 & Aug.16-$\star$1 \\
    2006 Aug 16 & 08:12 & 3.35 & 2.76 & 9.85 & 15.5 & 0.26 & 322 & +32 & 1.07 & Aug.16-$\star$2 \\
    2006 Aug 16 & 08:45 & 3.35 & 2.76 & 9.86 & 15.5 & 0.26 & 297 & +32 & 1.13 & Aug.16-$\star$2 \\    
    2006 Aug 16 & 09:00 & 3.35 & 2.76 & 9.86 & 15.5 & 0.26 & 285 & +32 & 1.17 & Aug.16-$\star$3 \\
    2006 Aug 16 & 09:18 & 3.35 & 2.76 & 9.86 & 15.5 & 0.26 & 272 & +32 & 1.23 & Aug.16-$\star$3 \\
    2007 Jul 12 & 13:15 & 3.31 & 2.69 & 9.78 & 15.5 & 0.26 & 211 & -38 & 1.03 & Jul.12-$\star$ \\
    2007 Nov 01 & 04:30 & 3.16 & 2.64 & 9.68 & 16.9 & 0.27 & 265 & -27 & 1.19 & Nov.01-$\star$ \\
    2007 Nov 01 & 06:06 & 3.16 & 2.64 & 9.68 & 16.9 & 0.27 & 191 & -27 & 1.12 & Nov.01-$\star$ \\
    \hline
  \end{tabular}

  \caption[Observation Conditions]{%
    Heliocentric distance ($\Delta$) and
    range to observer ($r$), 
    visual magnitude (V), 
    phase angle ($\alpha$), 
    angular diameter ($\phi$), and 
    Sub-Earth-Point (SEP) coordinates (longitude $\lambda$ and latitude
    $\varphi$)
    for each epoch (given in UT, the mid-observation time,
    here listed without light-time
    correction, although light-time corrections are included in
    all relevant computations in this paper).
    Airmass at the tabulated UT is 
    also reported. 
    The last column is a tag for the PSF used for
    deconvolution; see Table~\ref{tab-obs-psf} for a complete
    description of the stars.
    \label{tab-obs-condition}}

\end{center}
\end{table*}
%
%
%
\begin{table*}
\begin{center}

  \textbf{Observation Settings}\\
  \begin{tabular}{cccccccc}
    \hline\hline
    Date & Inst. & Filters & $\lambda_c$ & $\Delta \lambda$ & Images & $\Theta$ & ROI \\
    (UT)&   &       &  ($\mu$m)   &  ($\mu$m)        &   \#    & (km)     & (\%) \\
    \hline
    \hline
    2003 Oct 10$^{a}$ & NIRC2 & Kp & 2.124 & 0.35 & \textcolor{white}{0}4 & 57 & 60 \\
    2003 Oct 12$^{b}$ & NIRC2 & Kp & 2.124 & 0.35 & \textcolor{white}{0}9 & 57 & 60 \\
    2005 Feb 02$^{a}$ & NACO & J  & 1.265 & 0.25  & \textcolor{white}{0}8 & 37 & 60 \\
    2005 Feb 02$^{a}$ & NACO & H  & 1.66  & 0.33  & 12 & 48 & 55 \\
    2005 Feb 02$^{a}$ & NACO & Ks & 2.18  & 0.35  & 13 & 64 & 50 \\
    2005 Mar 12$^{a}$ & NACO & J  & 1.265 & 0.25  & \textcolor{white}{0}6 & 32 & 60 \\
    2005 Mar 12$^{a}$ & NACO & H  & 1.66  & 0.33  & \textcolor{white}{0}6 & 41 & 60 \\
    2005 Mar 12$^{a}$ & NACO & Ks & 2.18  & 0.35  & \textcolor{white}{0}5 & 54 & 60 \\
    2005 Mar 13$^{a}$ & NACO & J  & 1.265 & 0.25  & \textcolor{white}{0}6 & 32 & 60 \\
    2005 Mar 13$^{a}$ & NACO & H  & 1.66  & 0.33  & \textcolor{white}{0}6 & 41 & 60 \\
    2005 Mar 13$^{a}$ & NACO & Ks & 2.18  & 0.35  & \textcolor{white}{0}6 & 54 & 55 \\
    2005 May 08$^{c}$ & NACO & J  & 1.265 & 0.25  & \textcolor{white}{0}6 & 41 & 50 \\
    2005 May 08$^{c}$ & NACO & H  & 1.66  & 0.33  & \textcolor{white}{0}9 & 54 & 55 \\
    2005 May 08$^{c}$ & NACO & Ks & 2.18  & 0.35  & 13 & 70 & 50 \\
    2005 May 09$^{c}$ & NACO & H  & 1.66  & 0.33  & \textcolor{white}{0}9 & 54 & 60 \\
    2005 May 09$^{c}$ & NACO & Ks & 2.18  & 0.35  & \textcolor{white}{0}6 & 71 & 50 \\
    2006 Aug 16$^{d}$ & NIRC2 & Kp & 2.124 & 0.35 & 35 & 88 & 50 \\
    2007 Jul 12$^{e}$ & NIRC2 & Kp & 2.124 & 0.35 & \textcolor{white}{0}7 & 85 & 50 \\
    2007 Nov 01$^{e}$ & NIRC2 & Kp & 2.124 & 0.35 & 19 & 84 & 50 \\
    \hline
  \end{tabular}
  \caption[Observations Settings]{
    Observation settings, with filter characteristics
    (central wavelength $\lambda_c$ and bandwidth $\Delta\lambda$)
    for each camera (NIRC2 at Keck and NACO at VLT),
    number of images,
    theoretical size 
    ($\Theta$)
    of the resolution elements
    (estimated as $\lambda_c / D$, where $D$ is the
    diameter of the primary mirror),
    and the size of the Region of Interest (ROI) for each epoch (given in UT).
    Image scales are 0.010\arcsec/pix for NIRC2 and 0.013\arcsec/pix for NACO; both systems
    oversample the PSF for all wavebands.
    PIs for these observations were: $^{a}$C. Dumas,
    $^{b}$W. J. Merline, $^{c}$S. Erard, 
    $^{d}$J. D. Drummond, and $^{e}$A. Conrad.
    \label{tab-obs-settings}	
    \label{lasttable}
  }

\end{center}
\end{table*}
%
%
%
%
\begin{table*}
\begin{center}
  \textbf{Point-Spread-Function Observations}\\
  \begin{tabular}{ccccccccccc}
    \hline\hline
Name & Date & UT & Filter & Designation & RA & DEC & V & Airmass & \multicolumn{2}{c}{FWHM} \\
     & (UT) &    &      &    & (hh:mm:ss) & (dd:mm:ss) &(mag) & & (mas) & (km)\\
    \hline
    \hline
Oct.10-{\Large$\star$}1 & 2003 Oct 10 & 12:12 & Kp & HD 13093       & 02:07:47 & -15:20:46 &  8.70 & 1.27 & 78 & 102 \\
Oct.12-{\Large$\star$}1 & 2003 Oct 12 & 09:04 & Kp & HD 7662        & 01:16:26 & -12:31:50 & 10.35 & 1.25 & 56 &  73 \\
Oct.12-{\Large$\star$}2 & 2003 Oct 12 & 09:25 & Kp & HD 12628       & 02:03:25 & -17:01:59 &  8.17 & 1.39 & 52 &  68 \\
Feb.02-{\Large$\star$}1 & 2005 Feb 02 & 06:59 & J  & HD 109098      & 12:32:04 & -01:46:20 &  7.31 & 1.16 & 62 &  72 \\
Feb.02-{\Large$\star$}1 & 2005 Feb 02 & 06:56 & H  & HD 109098      & 12:32:04 & -01:46:20 &  7.31 & 1.16 & 62 &  72 \\
Feb.02-{\Large$\star$}1 & 2005 Feb 02 & 06:51 & Ks & HD 109098      & 12:32:04 & -01:46:20 &  7.31 & 1.16 & 64 &  74 \\
Feb.02-{\Large$\star$}2 & 2005 Feb 02 & 08:30 & J  & HD 109098      & 12:32:04 & -01:46:20 &  7.31 & 1.08 & 62 &  71 \\
Feb.02-{\Large$\star$}2 & 2005 Feb 02 & 08:27 & H  & HD 109098      & 12:32:04 & -01:46:20 &  7.31 & 1.08 & 64 &  74 \\
Feb.02-{\Large$\star$}2 & 2005 Feb 02 & 08:24 & Ks & HD 109098      & 12:32:04 & -01:46:20 &  7.31 & 1.08 & 64 &  75 \\
Mar.12-{\Large$\star$}  & 2005 Mar 12 & 06:28 & J  & HD 109098      & 12:32:04 & -01:46:20 &  7.31 & 1.10 & 74 &  73 \\
Mar.12-{\Large$\star$}  & 2005 Mar 12 & 06:25 & H  & HD 109098      & 12:32:04 & -01:46:20 &  7.31 & 1.10 & 64 &  63 \\
Mar.12-{\Large$\star$}  & 2005 Mar 12 & 06:21 & Ks & HD 109098      & 12:32:04 & -01:46:20 &  7.31 & 1.10 & 64 &  63 \\

Mar.13-{\Large$\star$}  & 2005 Mar 13 & 05:02 & J  & HD 109098      & 12:32:04 & -01:46:20 &  7.31 & 1.11 & 68 &  67 \\
Mar.13-{\Large$\star$}  & 2005 Mar 13 & 05:04 & H  & HD 109098      & 12:32:04 & -01:46:20 &  7.31 & 1.11 & 58 &  57 \\
Mar.13-{\Large$\star$}  & 2005 Mar 13 & 05:07 & Ks & HD 109098      & 12:32:04 & -01:46:20 &  7.31 & 1.11 & 53 &  52 \\
May.08-{\Large$\star$}1 & 2005 May 08 & 22:51 & H  & NGC 2818 TCW E & 09:15:50 & -36:32:36 & 12.21 & 1.02 & 51 &  65 \\
May.08-{\Large$\star$}1 & 2005 May 08 & 22:47 & Ks & NGC 2818 TCW E & 09:15:50 & -36:32:36 & 12.21 & 1.02 & 40 &  50 \\
May.08-{\Large$\star$}2 & 2005 May 09 & 01:58 & J  & BD+20 2680     & 12:05:53 & +19:26:52 & 10.13 & 1.39 &113 & 145 \\
May.08-{\Large$\star$}2 & 2005 May 09 & 01:52 & H  & BD+20 2680     & 12:05:53 & +19:26:52 & 10.13 & 1.39 & 67 &  86 \\
May.08-{\Large$\star$}2 & 2005 May 09 & 01:47 & Ks & BD+20 2680     & 12:05:53 & +19:26:52 & 10.13 & 1.39 & 64 &  82 \\
May.08-{\Large$\star$}3 & 2005 May 09 & 03:17 & J  & BD-06 4131     & 15:05:39 & -06:35:26 & 10.33 & 1.09 & 79 & 101 \\
May.08-{\Large$\star$}3 & 2005 May 09 & 03:26 & H  & BD-06 4131     & 15:05:39 & -06:35:26 & 10.33 & 1.09 & 66 &  84 \\
May.08-{\Large$\star$}3 & 2005 May 09 & 03:38 & Ks & BD-06 4131     & 15:05:39 & -06:35:26 & 10.33 & 1.09 & 64 &  82 \\
May.09-{\Large$\star$}1 & 2005 May 10 & 00:27 & H  & BD+20 2680     & 12:05:53 & +19:26:52 & 10.13 & 1.47 & 71 &  92 \\
May.09-{\Large$\star$}1 & 2005 May 10 & 00:05 & Ks & BD+20 2680     & 12:05:53 & +19:26:52 & 10.13 & 1.55 & 62 &  80 \\
May.09-{\Large$\star$}2 & 2005 May 10 & 01:45 & H  & BD-06 4131     & 15:05:39 & -06:35:26 & 10.33 & 1.40 & 72 &  93 \\
May.09-{\Large$\star$}2 & 2005 May 10 & 01:56 & Ks & BD-06 4131     & 15:05:39 & -06:35:26 & 10.33 & 1.34 & 60 &  78 \\

May.09-{\Large$\star$}3 & 2005 May 10 & 08:28 & H  & BD-06 4131     & 15:05:39 & -06:35:26 & 10.33 & 1.92 & 66 &  85 \\
May.09-{\Large$\star$}3 & 2005 May 10 & 08:38 & Ks & BD-06 4131     & 15:05:39 & -06:35:26 & 10.33 & 2.05 & 63 &  82 \\

Aug.16-{\Large$\star$}1 & 2006 Aug 16 & 07:12 & Kp & NLTT 45848     & 18:03:01 & +17:16:35 &  9.89 & 1.01 & 43 &  86 \\
Aug.16-{\Large$\star$}2 & 2006 Aug 16 & 08:15 & Kp & NLTT 45848     & 18:03:01 & +17:16:35 &  9.89 & 1.07 & 42 &  83 \\
Aug.16-{\Large$\star$}3 & 2006 Aug 16 & 09:22 & Kp & NLTT 45848     & 18:03:01 & +17:16:35 &  9.89 & 1.25 & 42 &  83 \\
Aug.16-{\Large$\star$}4 & 2006 Aug 16 & 10:27 & Kp & NLTT 45848     & 18:03:01 & +17:16:35 &  9.89 & 1.63 & 42 &  84 \\

Jul.12-{\Large$\star$}  & 2007 Jul 12 & 13:10 & Kp & G 27-28        & 22:26:34 & +04:36:35 &  9.73 & 1.04 & 39 &  76 \\

Nov.01-{\Large$\star$}  & 2007 Nov 01 & 04:12 & Kp & HD 214425      & 22:38:07 & -02:53:55 &  8.28 & 1.28 & 44 &  84 \\
    \hline
  \end{tabular}

  \caption[Point Spread Function Observations]{%
    Basic information (designation, coordinates, and
    visual magnitude) for all PSF stars observed (see 
    Table~\ref{tab-obs-condition}). 
    The time and
    airmass of these observations are given.
    We also report the measured FWHM for each star and
    each filter
    in milliarcsec and in km at Pallas'
    distance, thus
    giving an idea of the AO correction achieved at that time.
    \label{tab-obs-psf}	
    \label{lasttable}
  }

\end{center}
\end{table*}
%
%
  \section{Data reduction}
\begin{figure}
\begin{center}

  \textbf{Pallas in K-band}\\
  \resizebox{0.9\hsize}{!}{\includegraphics{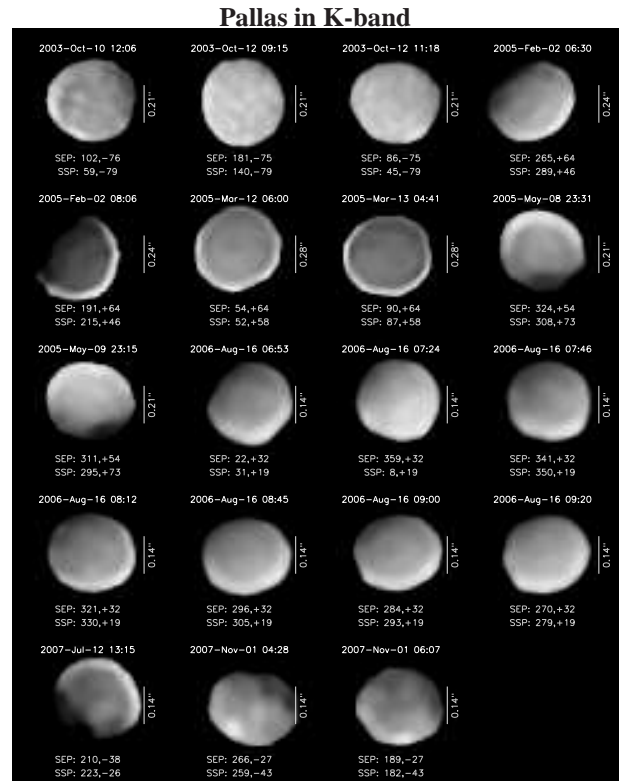}}

    \caption[Views of Pallas]{
      Selected views of (2) Pallas in \textsl{K}-band, oriented
      to have the rotation axes aligned vertically, with the determined
      spin vector directed toward the top.
      The values of the Sub-Earth-Point longitude (SEP$_\lambda$), measured
      positively from 0 to 360 degrees in a right-hand system \citep[following
      IAU recommendations:][]{2007-CeMDA-98-Seidelmann}, and 
      Sub-Earth-Point latitude (SEP$_\varphi$) are
      indicated under each view.
      The Sub-Solar-Point (SSP)
      [longitude, latitude] is also shown.
      Arbitrary brightness and contrast cuts were applied to highlight the surface
      features.
      The edge-brightness
      ringing present in some images is an artifact (see
      section~\ref{sssec-contour}) from
      deconvolution \citep[see][for further
        explanation]{2008-AA-478-Carry}.
      The impact of
      these artifacts is
      limited in the current study: 
      it does not influence the shape and we
      discard the perimeter of the asteroid in each view
      for 
      the albedo map (see section~\ref{sec-surface}).
      From these images, the irregular shape of Pallas, as well as important
      brightness variations across the
      surface, can be detected easily.
    }

    \label{fig-views_of_pallas}
    \label{lastfig}

  \end{center}
\end{figure}

    \indent We reduced the data using 
    standard techniques
    for
    near-infrared images. A bad pixel mask was made by combining the
    hot and dead pixels found from the dark and flat-field frames. The
    bad pixels in our calibration and science images were then
    corrected by replacing their values with the median of the
    neighboring pixels (7$\times$7 pixel box). Our sky frames were obtained
    from the median of each series of
    dithered science images, and then
    subtracted from the corresponding science images to remove
    the sky and instrumental background. By doing so, the dark current
    was also removed. Finally, each image was divided by a normalized
    flat-field to correct the pixel-to-pixel sensitivity
    differences of the detector. \\
    \indent We then restored the images to optimal angular-resolution by
    applying the \textsc{Mistral} deconvolution algorithm 
    \citep{these-fusco, 2004-JOSAA-21-Mugnier}. This image restoration algorithm 
    is particularly well suited to
    deconvolution of objects with sharp edges, such as
    asteroids. Image
    restoration techniques are known to be
    constrained by the limitation of 
    trying to measure/estimate the precise instrumental
    plus
    atmospheric responses at the exact time of
    the science observations.
    \textsc{Mistral} is an iterative myopic deconvolution method, which estimates
    both the most probable object, and the PSF, from analysis of science
    and reference-star images \citep[see][for details]{2004-JOSAA-21-Mugnier}. 
    In total, we obtained 186 images of Pallas with a
    spatial resolution (Table~\ref{tab-obs-settings})
    corresponding
    to the diffraction limit of the telescope
    (which we estimate by 
    $\lambda$/$D$, with $\lambda$
    the wavelength and $D$
    the telescope diameter) and the range of the observer given in
    Table~\ref{tab-obs-condition}.
    A subset of the restored images is presented
    in Fig.~\ref{fig-views_of_pallas}.
%
%
%
%
%

\section{Size, shape and spin-vector coordinates}
  \indent Disk-resolved observations
  (from space, ground-based AO, radar, or occultations)
  provide strong constraints on asteroid shape.
  The limb contour recorded is a direct
  measurement of the asteroid's outline on the sky.
  Combination of such contours leads to the
  construction of an asteroid shape
  model and an associated
  pole solution \citep{2007-Icarus-191-Conrad}.
  To improve our shape model, we combined our AO data
  with the numerous light-curves
  available for Pallas 
  (51 of them, which led \citet{2003-Icarus-164-Torppa} to their own
    shape model).
  \subsection{Discrimination of the pole solution}
\begin{figure}
\begin{center}

  \textbf{Pole Solution Selection}\\
  \resizebox{0.9\hsize}{!}{\includegraphics{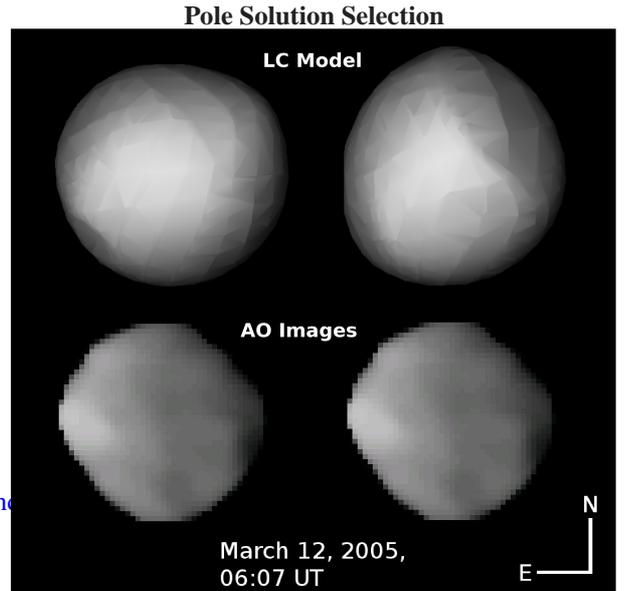}}\\

    \caption[Pole Solution Selection]{
      Plane-of-sky synthetic images of the two light-curve shape models
      compared with AO images for the epoch 2005 Mar 12 UT
      (in J-band at the VLT).
      Because the light-curve shape models have no absolute dimensions,
      their size with respect to the Pallas images is not relevant
      here and we focus on the overall shape only.
      The AO images are identical left and right; they are repeated for
      ease of comparison.
      This comparison is used only to discriminate between
      the two pole solutions.
      Because the
      shape model on the right, largely asymmetric in this view,
      does not reproduce correctly
      the shape of Pallas,  
      we reject it in favor of
      the other (left) model instead
      (see text).
    }
    \label{fig-wrong-pole}
    \label{lastfig}

  \end{center}
\end{figure}

    \indent Due to an ambiguity inherent in the method and observation
    geometry, it is sometimes impossible to discriminate between the two
    possible pole
    solutions obtained from the light-curve inversion process.
    Therefore, we
    produced the two contours of both
    (light-curve-derived) shape models, as
    projected onto the plane of the sky for the time of
      our AO observations,
    and compared them 
    with our images of Pallas,
    as shown in Fig.~\ref{fig-wrong-pole}.
    This simple comparison
    \citep{2003-Icarus-162-Cellino,2006-Icarus-185-Marchis} allowed us to
    reject one pole solution (Fig.~\ref{fig-wrong-pole}, right) in
    favor of the other (Fig.~\ref{fig-wrong-pole}, left)
    based on its
    poor representation of the asteroid contour.
    Even though
    the selected pole solution and its associated shape model
      rendered better the AO images,
    it was clear that
    the shape
    model still 
    needed improvement.
    Indeed, the light-curve inversion algorithm
    \citep{2001-Icarus-153-Kaasalainen-a}
    associates photometric variation with shape and not with
    albedo. The presence of albedo
    markings (as found here, see section \ref{sec-surface}) would
    thus lead to an 
    erroneous shape.
    We discuss the development of
    a new shape model in
    section~\ref{ssec:shape}.\\
    \indent Once we had rejected one of the possible
    pole-solution regions (from light-curves, above), we refined the pole solution
    by fitting (next section) against our ensemble of AO images.
    We find the 
    spin-vector coordinates
    of Pallas to be
    within
    5\ensuremath{^\circ}~of arc 
    of [$\lambda = 30$\ensuremath{^\circ},
      $\beta = -16$\ensuremath{^\circ}] in the Ecliptic J2000.0 
    reference frame (Table~\ref{tab-PoleSolution}).
    This value is roughly in agreement with the value
    (40\degr, -16\degr) in Ecliptic B1950 coordinates
    [equivalent to
    (41\degr, -16\degr) in Ecliptic J2000]
    found by
    \citet{2007-Icarus-192-Kryszczynska} from a synthesis
    of reported pole solutions (mainly from indirect
    methods).
%

    Recent pole solutions are reported near our
    solution.
    \citet{2008-Icarus-197-Drummond} give a solution
    of (32\ensuremath{^\circ},-21\ensuremath{^\circ}) $\pm$
    6\ensuremath{^\circ}~(in Ecliptic J2000)
    based on AO observations done at
    Lick Observatory. In a follow-up report by the same authors
    \citep{2009-Icarus-Drummond}, from AO observations at Keck
    (included as a subset here), a solution of
    (34\ensuremath{^\circ},-27\ensuremath{^\circ})
    $\pm$ 3\ensuremath{^\circ} is quoted.
    These solutions are in rough agreement with the
    value derived here, and we note that 
    \citet{2009-Icarus-Drummond}
    list their errors to be model-fit errors only, essentially
    precisions of the measures, while indicating there may be
    systematic errors that are not included in the quoted
    error.  Because our solution is derived from a larger number
    of epochs, and because it also considers extensive light-curve
    datasets, we think the difference from the solution
    of \citet{2009-Icarus-Drummond}
    is the result of systematic errors in their more limited
    dataset.\\
%
%
%
%
%
    \indent The present pole solution implies a high obliquity
    of $\sim$84\ensuremath{^\circ},
    which means seasons on Pallas have high contrast.
    Large portions of both hemispheres will experience 
    extended periods of constant sunlight or constant darkness over
    Pallas' orbital period of 4.6 years.  Locations near the poles 
    would remain in total sunlight or darkness for as long as 
    two years.\\
%
%
%
%
%
%
\begin{table*}
\begin{center}

  \textbf{Pole solution}\\
  \begin{tabular}{ccccc}
    \hline\hline
    $P_s$ & Ecliptic & Equatorial & $t_0$ & $W_0$\\
    (h)  & ($\lambda_0$, $\beta_0$ in \degr) &
           ($\alpha_0$, $\delta_0$ in \degr) &
    (JD) & (\degr) \\
    \hline
    7.8132214 $\pm$ 0.000002 & (30, -16) $\pm$ 5 &
                              (33, -3)  $\pm$ 5 &
    2433827.77154 
    & 38 $\pm$ 2 \\
    \hline
  \end{tabular}
  \caption[Pole Solution]{
    Sidereal
    period ($P_s$, determined from the 51 light-curves) of Pallas
    and 
    spin-vector coordinates in
    Ecliptic J2000.0 ($\lambda_0$, $\beta_0$)
    and Equatorial
    J2000.0 ($\alpha_0$, $\delta_0$) reference frames
    and the reference epoch $t_0$ (see section~\ref{subsec-pallas_model}).
    We also report the rotational phase ($W_0$) at epoch J2000.0,
    following IAU guidelines \citep{2007-CeMDA-98-Seidelmann}.
    The rotational phase $W$ of Pallas at any time is then given
      by $W = W_0 + \dot{W} \times d$, where 
      $d$ is the number of days since epoch J2000.0 and $\dot{W}$ is
      Pallas rotation rate, 1105.8036\degr/day.
    \label{tab-PoleSolution}
    \label{lasttable}}

\end{center}
\end{table*}
%
%
  \subsection{Construction of the shape model\label{ssec:shape}}
    \indent We constructed a shape model, based on both AO
    observations and optical light-curves, to render the aspect of
    Pallas at each epoch.

    \subsubsection{Contour measurement\label{sssec-contour}}
\begin{figure}
\begin{center}

  \textbf{Contour extraction}\\
  \resizebox{0.9\hsize}{!}{\includegraphics{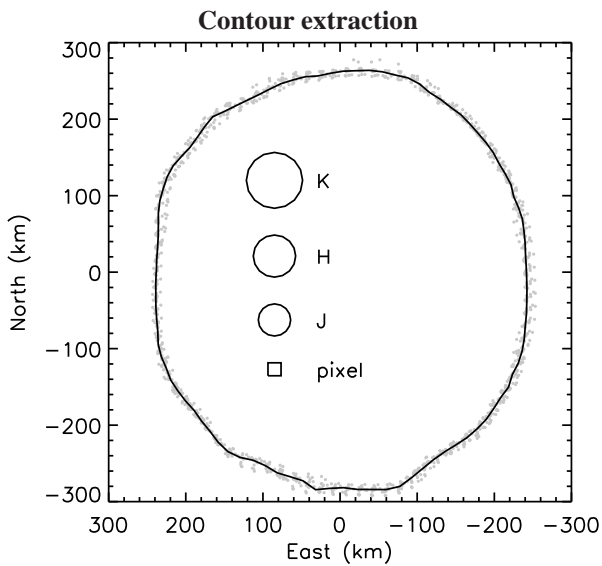}}

    \caption[Contour extraction]{
      Example of contour extraction for the observations at the epoch 2005 May 8,
      from the VLT (see
      Tables~\ref{tab-obs-condition}~\&~\ref{tab-obs-settings}), with 
      geometry SEP$_\lambda$=307\degr,
      SEP$_\beta$=+54\degr.
      North is up and east is to the left.
      Axes
      are in kilometers (given
      Pallas' distance and angular size at observing time).
      The final contour (dark line) was obtained by taking the median of 13
      individual edge measurements (grey spots).
      One can verify the
      consistency between each individual contour ($\sigma\sim$13 km).
      Similar composite contours were created for each epoch listed in 
      Table~\ref{tab-obs-condition} and used
      to constrain the shape of Pallas during light-curve inversion (see text).
      The sizes of the J, H, and K resolution elements, as well as
      the pixel size, are also shown.
    }
      \label{fig-shape}
    \label{lastfig}

  \end{center}
\end{figure}

      \indent The deconvolution process is an ill-posed inverse problem
      \citep{1974-Tikhonov} and can
      introduce artifacts in the restored images.
      Although we carefully cross-checked the images after the deconvolution
      process, the presence of artifacts was still possible.
      Because reliable information is
      provided by limb contours, which are far less subject to
      artifact contamination in the deconvolution
      process \citep[see][Fig.~2]{2006-Icarus-185-Marchis}, we
      chose to discard the albedo information from our images at this stage. \\
      \indent We measured 186 limb contours using
      the Laplacian of a Gaussian
      wavelet transform \citep{2008-AA-478-Carry} of the Pallas frames.
      Then, to minimize introduction of artifacts,
      we took the median contour (Fig.~\ref{fig-shape})
      of each epoch (Table~\ref{tab-obs-condition})
      and used them as fiducials during the light-curve inversion.
      For each observational series, all frames were taken 
      within a span of 4--5 minutes, during which Pallas rotated only
      about 3--4\ensuremath{^\circ}. This translates into a
      degradation of the spatial information that is
      much lower than the highest angular resolution
      achieved in our images. \\

%
%
%
%
%
    \subsubsection{KOALA}
      \indent The shape and spin
      model was created by combining the two data modes, photometry
      (light-curves) and 
      adaptive-optics contours, with the general principle described in
      \citet{2006-IP-22-Kaasalainen}: the joint chi-square is minimized with the
      condition that the separate chi-squares for the two modes be
      acceptable
     (the light-curve fit deviation is 0.019 mag and the profile fit
      deviation is 0.4 pixel).
      The light-curve fitting procedure is described in
      \citet{2001-Icarus-153-Kaasalainen-b},
      and the edge fitting method and the choice of weights for
      different data modes is described in detail in
      Kaasalainen [submitted to Inverse Problems and Imaging].
      We also used a smoothness
      constraint (regularizing function) to prevent artificial details in the
      model, \textsl{i.e.}, we chose the simplest model
      that was capable of fitting successfully the data.
      Since Pallas is a rather regular body
      to a first approximation, and the
      data resolution is limited, we chose to use a function series
      in spherical harmonics to represent the radii lengths in fixed directions
      \citep[\textsl{see}][]{2001-Icarus-153-Kaasalainen-a}.
      In addition to reducing the number of
      free parameters and providing global continuity, the function series, once
      determined, gives a representation that can be directly evaluated for any
      number of radii (or any tessellation scheme) chosen without having to
      carry out the inversion again. The number of function coefficients, rather
      than the tessellation density, determines the level of resolution.\\
      \indent As discussed in Kaasalainen [submitted to Inverse Problems and Imaging],
      profile and shadow edges (when several
      viewing angles are available) contain, in fact, almost as much information
      on the shape and spin as direct images. In our case, the edges are also
      considerably more reliable than the information across the
      deconvolved disk
      (see section \ref{sssec-contour}),
      so the modeling
      is indeed best done by combining edges, rather than images, with
      light-curves. The procedure is directly applicable to combining photometry
      and occultation measurements as well.
      The technique of combining these three data modes we call
      KOALA for Knitted Occultation,
      Adaptive optics, and Light-curve Analysis.
  \subsection{The irregular shape of Pallas\label{subsec-pallas_model}}
\begin{figure}
\begin{center}
\vspace*{-1cm}
  \textbf{Pallas Shape Model}\\
  \resizebox{0.6\hsize}{!}{\includegraphics{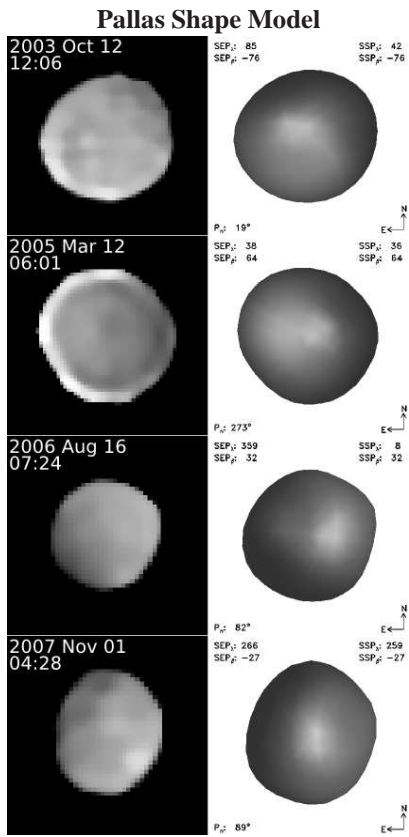}}

    \caption[Pallas Shape Model]{
      Comparison of the
      shape model derived in this study (right)
      with the original frames (left).
      Sub-Earth-Point (SEP) and Sub-Solar-Point (SSP) coordinates as
      well as the 
      pole angle $p_{n}$
      (defined as the angle in the plane of the sky
      between the celestial north and the
      projected asteroid spin-vector, measured counter-clockwise,
      from north through east) are labeled for
      each representation of the model.
      These representations were obtained using the Eproc ephemeris generator
      \citep{1998-IMCCE-Berthier}.
      The improvement in the shape model can be
      seen by comparing the
      second epoch presented here (2005 Mar 12) with the shape model
      presented in the upper-left panel of
      Fig.~\ref{fig-wrong-pole}.
    }
    \label{fig-model}
    \label{lastfig}
  \end{center}
\end{figure}

    \indent From the combination of light-curves and
    high angular-resolution images, we found Pallas to be an irregular
    asteroid  with significant departures from an
    ellipsoid, as visible in Fig.~\ref{fig-views_of_pallas}.
    Our shape model, presented in Fig.~\ref{fig-model},
    is available
    either on request\footnotemark[1]
    or from the
    Internet\footnotemark[2].
    Useful parameters (coordinates of the SEP and SSP as well as
    pole angle)
    to display the shape model of Pallas, as seen on the plane of the sky
    at any time,
    can be computed 
    from the values reported in
    Table~\ref{tab-PoleSolution} and
    the following equation from
    \citet{2001-Icarus-153-Kaasalainen-b}, which transforms
    vectors (\textsl{e.g.} Earth-asteroid vector)
    from the ecliptic reference frame ($\vec{r}_{ecl}$) into
    the reference frame of the shape model ($\vec{r}_{ast}$).
\begin{equation}      
  \vec{r}_{ast} = \mathcal{R}_z \left(\frac{2\pi}{P_s} (t - t_o)\right)
                  \mathcal{R}_y \left(\frac{\pi}{2} - \beta_0 \right) 
                  \mathcal{R}_z(\lambda_0) \vec{r}_{ecl}
\end{equation}
    where $\lambda_0$, $\beta_0$ are the pole coordinates in the
    Ecliptic reference frame,
    $P_s$ the sidereal period (Table~\ref{tab-PoleSolution}),
    $t_0$ the epoch of reference
    (chosen arbitrarily as $t_0 = 2433827.77154$ JD, the starting time of the first
    light-curve used here), and
    $t$ is the time.
    $\mathcal{R}_i(\alpha)$ is the rotation matrix representing a
    rotation by angle $\alpha$ about axis $i$, in the positive sense.
    Then, we report in Table~\ref{tab-triaxial}
    our best-fit tri-axial ellipsoid
    values, with measurement dispersion, compared with the
    \citet{2009-Icarus-Drummond}
    and \citet{2009-LPI-Schmidt}
    studies.\\

\footnotetext[1]{BC:\href{mailto:benoit.carry@obspm.fr}{benoit.carry@obspm.fr} or 
                 MK:\href{mailto:mjk@rni.helsinki.fi}{mjk@rni.helsinki.fi}}
\footnotetext[2]{DAMIT:\href{http://astro.troja.mff.cuni.cz/~projects/asteroids3D/}
                 {http://astro.troja.mff.cuni.cz/$\sim$projects/asteroids3D/}}
%
%
%
%
%
%
\begin{table*}
\begin{center}

  \textbf{Tri-Axial Solution}\\
  \begin{tabular}{cccccccc}
    \hline\hline
    & $a$ & $b$ & $c$ & $R$ & $a$/$b$ & $b$/$c$ & $V$  \\
    & (km)&(km)&(km)&(km)& & &($\times 10^6$ km$^{3}$) \\
    \hline
    This work  & 275 & 258 & 238  & 256 & 1.06 & 1.09 & 70 \\
    1\thinspace$\sigma$ error    &  4  &  3  &  3   &  3  & 0.03 & 0.03 &  3 \\
    \hline
     \citet{2009-Icarus-Drummond} & 274 & 252 & 230 & 251 & 1.09 & 1.10 &  66 \\
    1\thinspace$\sigma$ error  &  2  &  2  &  7  &  3  & 0.01 & 0.03 &   2 \\
    \hline
    \citet{2009-LPI-Schmidt}  & 291 & 278 & 250 & 272 & 1.05 & 1.11 &  85 \\
    1\thinspace$\sigma$ error   &  9  &  9  &  9  &  9  & 0.06 & 0.08 &   8 \\
    \hline
  \end{tabular}

  \caption[Tri-Axial Solution]{
     Best-fit solutions for tri-axial radii ($a$, $b$, $c$) for
     Pallas.  Also given are the
    mean radius $R=\sqrt[3]{abc}$, 
    axial ratios, and 
    volume ($V$),
    with their uncertainties
    for this work, and for two other recent determinations:
    Keck \citep{2009-Icarus-Drummond}
    and
    HST \citep{2009-LPI-Schmidt}.
    For the current study,
    the dimension uncertainties are derived
    by scaling the shape model, for 
    each image, to best render the
    asteroid contour. The dimensions are then determined using the
    mean values of the scaling factor for the ensemble of images, 
    and the 1\thinspace$\sigma$ reported
    here comes from the standard deviation of the population of
    scaling factors.
    \label{tab-triaxial}
    \label{lasttable}
  }

\end{center}
\end{table*}
%

\begin{figure}
\begin{center}

\vspace*{-1cm}
  \textbf{Comparison with occultations}\\
  \includegraphics[width=.5\textwidth]{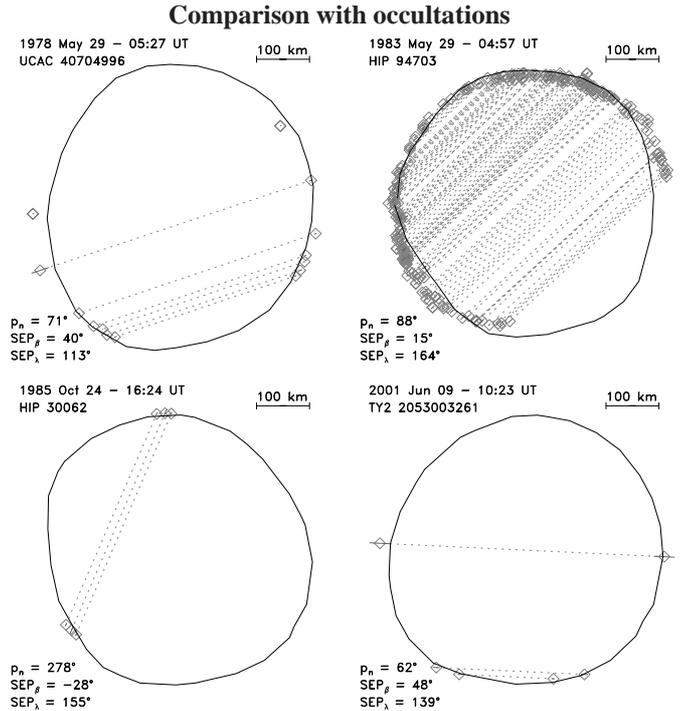}%

  \caption[Comparison with occultations]{
    Comparison of our shape model
    with occultation chords 
    for four occultation events.
    We use the method described by
    \citet{1999-IMCCE-Berthier}
    to convert the occultation timings reported by
    \citet{2008-Occultations} to their print on the plane of the sky.
    Celestial north is up and east is to the left.
    For each occultation, we list the date and time (UT), the occulted
    star, and
    the SEP coordinates and pole angle
    $p_n$ (defined in Fig.~\ref{fig-model}) for Pallas.
    \add{For each chord, the 
      diamonds represent the exact time of disappearance and
      reappearance as reported by the observer, linked by the dashed
      lines, and the solid lines represent the error stated
      by the observer.}
    \label{fig-occ}
    \label{lastfig}
  }
  \end{center}
\end{figure}

\begin{figure}[!p]
\begin{center}

\vspace*{-1cm}
  \textbf{Occultation radial profiles}\\
  \includegraphics[width=.5\textwidth]{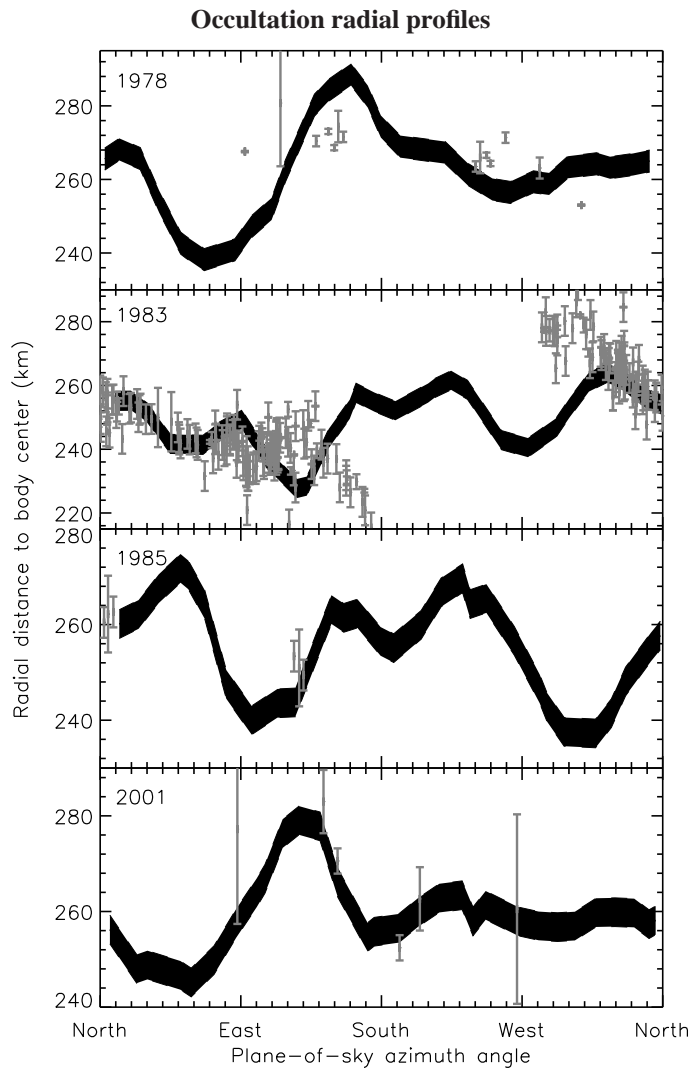}%

  \caption[Occultation radial profiles]{%
    \add{
    Higher resolution view of the contours
    for the four occultations presented in
    Fig.~\ref{fig-occ}. 
    The distance to the body center of
    the shape model, and the locations of the chord endpoints, are
    represented
    as a function of an angle in the plane of sky
    (measured counter-clockwise from the celestial north through east). 
    The chord endpoints and their associated errors are represented by
    the grey bars. The black bands show the contour of the
    shape model with its uncertainty.}
    \label{fig-occdev}
    \label{lastfig}
  }
  \end{center}
\end{figure}

    \indent The dimensions derived here for Pallas
     are somewhat larger, at the few-$\sigma$ level, than
    those derived by \citet{2009-Icarus-Drummond}.
    The quoted errors by Drummond et al., however, do not
    include possible systematic effects, which they indicate
    could be in the range of 1-2\% of the values.  Once their
    quoted errors are augmented to include systematics, their
    dimensions are entirely consistent with our derived values.
    The smaller error bar quoted here
    for the $c$ dimension results from
    more observations, taken over a wider span of
    SEP latitudes (Table~\ref{tab-obs-condition}).
    We continue to refine our estimates of our absolute accuracy,
    but we are confident it is significantly 
    smaller than the difference (16 km) between our value
    for the mean radius ($R = 256 \pm 3$ km)
    and that from HST WFPC2 ($R=272 \pm 9$ km) of
    \citet{2009-LPI-Schmidt}.
    Our method to determine the error relies on searching for the
    minimum and maximum possible dimensions of the shape-model
    contours that would
    be consistent with the images.  Therefore, the quoted errors are the
    best approximation to absolute accuracy at this time.  Further, in
    search of possible systematics in our technique, we have run a
    range of simulations for Pallas and a few other asteroids for which
    we have data.  The preliminary results are that the errors quoted
    here appear to include systematics and, in any case, the absolute
    errors are unlikely to be much larger than the error quoted here.
    Two issues that may be relevant to systematics of the HST
    observations relative to ours, 
    are 
    \add{1) the WFPC2 PSF, although stable and well characterized,
      is under-sampled} (giving a resolution set by
    2 pixels, or about 149 km at all wavebands)
    and
    \add{2)} the lack of
    deconvolution (for 
    size determination),
    which would naturally result in larger values
    \add{\citep[see Fig. 3 in][for
        instance]{2006-Icarus-185-Marchis}}. \\  
%
    \add{
    \indent Next, we present in Fig.~\ref{fig-occ}
    our shape model, oriented on the sky to correspond to the times of
    four stellar occultations by Pallas
    \citep{2008-Occultations}.
    To assess quantitatively the match between the shape model
    from the AO/light-curve observations and the occultation chords, we 
    show in Fig.~\ref{fig-occdev} the radius of the shape model as a function of
    azimuth angle from the center of the projected figure of the
    body, along with the measured endpoints of the chords (and their
    associated uncertainties). 
    The correspondence for the 1985 and
    2001 events is quite good, while that for 1978 and 1983 is less so.
    In Table~\ref{tab-occ}, we display the RMS deviation of the chords from the
    shape model, both in km and in terms of the occultation
    uncertainties ($\sigma_c$). 
    The 1985 and 2001 events are about 1\thinspace$\sigma$, even without
    making an attempt to modify the shape model. The 3\thinspace$\sigma$ for
    1983 is still good, considering that this is not a description
    of a fit of a model to these (occultation) data, but instead 
    assessing how well a model, fit to other data sets, corresponds
    to the occultation data.
    The 6\thinspace$\sigma$ deviation for 1978 is driven
    almost entirely by one chord
    (the RMS deviation without taking
    this chord into account drops to 1.3\thinspace$\sigma$, see
    Table~\ref{tab-occ}).}\\
    \indent \add{Consideration of the RMS deviation
      in km shows that the 
      observed deviation between the shape model and the occultation
      chords is on the same scale as
      possible topographic features. The 
      localized deviations observed may thus reflect
      the presence of local topography.
      We found the RMS deviation similar to the 
      uncertainty resulting from the typical 
      occultation-timing error of 0.3 s.
      Here, we are demonstrating the potential utility of KOALA, by
      showing rough quantitative agreement between occultation and
      AO/light-curve-derived shape models. We are showing that
      the occultations are consistent with our triaxial dimensions
      as well as the general shape.  For future work, we will use the
      occultation data as additional constraints on the shape model
      itself, modifying the shape accordingly}.\\
    \add{
      \indent Although the occultations show that the shape model
      is globally correct, there may exist
      local topography where
      no limb measurements were available to constrain the shape.
      As an exemple, t}he
    flat region or facet at the S-E limb
    \add{in the upper right of Fig.~\ref{fig-occ}} is not
    borne out by the chords \add{($\sim$ 20 km mismatch between the
      model and the occultation chords)}.
    This facet appears in the northern hemisphere of
    Pallas,
    where a large dark albedo patch is also present (see
    section~\ref{subsec-mapping-description}).
    Because our technique does not yet take into account the albedo information
    during the inversion, a dark patch may be
    misrepresented as a deficit/depression if no limb measurement from AO
    constrains it.
    This highlights the need for future development of the KOALA
    technique, including the effects of albedo, 
    and the need for continued acquisition of high-quality 
     imaging at the widest range of geometries (SEP longitude and latitude).
%
%
%
%

\begin{table}
\begin{center}

  \textbf{Comparison of the Occultations and the Shape Model}\\
  \begin{tabular}{ccccc}
    \hline\hline
    Occultation Date & Chords & $<\sigma_c>$ & RMS & RMS \\
    (UT) & (\#) & (km) & (km) & ($\sigma_c$) \\
    \hline
    1978 May 29 - 05:27 $^{\alpha}$ &  7  &  5.4 & 12.9 & 6.1  \\  
    1978 May 29 - 05:27 $^{\beta}$  &  6  &  6.2 &  6.3 & 1.3  \\  
    1983 May 29 - 04:57             & 130 &  7.6 &  9.6 & 3.7  \\
    1985 Oct 24 - 16:24             &  3  &  9.6 &  6.1 & 0.8  \\
    2001 Jun 09 - 10:23             &  3  & 19.4 &  7.5 & 0.6  \\
    \hline
    \multicolumn{4}{l}{$\alpha$ \footnotesize{including the KAO chord}}\\
    \multicolumn{4}{l}{$\beta$ \footnotesize{without the KAO chord, with figure-center readjusted
      accordingly}}
  \end{tabular}
  \caption[Comparison of the Occultations and the Shape Model]{%
    \add{
    For each occultation event, 
    we give the RMS deviation of the
    endpoints of a chord from the shape-model edge (defined as the
    intersection of the shape-model edge with a line from the
    body center to the chord endpoint).  The values are shown in
    km and also computed in terms of the
    uncertainties, $\sigma_c$, of the
    chord endpoints (computed individually for each endpoint,
    because they vary).
    Thus, the last column shows how far from the 
    edge the chord endpoints typically lie, in terms of chord
    uncertainty.
    The column labeled
    $<$$\sigma_c$$>$ is the average value of the 
    chord uncertainty for that epoch, reported in km.
    Considering these values are
    not representing a goodness of fit, because the model is not
    being adjusted for these values, the deviations are reasonable
    for all but 1978.
    For 1978, the RMS deviation is dominated by
    one chord, from Kuiper Airborne Observatory (KAO),
    and is improved substantially by removal of this
    one chord (line labeled $\beta$).
    }
  \label{tab-occ}
  }
\end{center}
\end{table}

%
%
%
    \indent Because of the high inclination of its
    orbit (35\degr), Pallas remains
    above or below the canonical
    asteroid ``belt'' (and the ecliptic plane)
    most of the time. As a result, 
    mass determinations for Pallas generally
    show poor agreement,
    depending on the method used
    \citep[see][for a review]{2002-AsteroidsIII-2.2-Hilton}:
    perturbation of Mars
    \citep[e.g.,][]{1989-Icarus-80-Standish},
    asteroid close encounters \citep[\textsl{e.g.,}][]{2001-AA-365-Goffin}, or
    ephemeris theory \citep{2008-AA-477-Fienga}.
    To be conservative, we
    use the conservative estimate from
    \citet{2000-AA-360-Michalak}, which includes
    consideration of most previous mass estimates for Pallas.
    That value is
    1.2 $\times$
    10$^{-10}$ M$_\odot$, with an uncertainty of $\pm$ 0.3 $\times$
    10$^{-10}$ M$_\odot$.
    Combined with our new estimates for Pallas' dimensions, and
    hence volume, we derive a density for Pallas of
     $\rho = 3.4 \pm$ 0.9 g.cm$^{-3}$, 
    where the uncertainty on the mass now dominates the
    density uncertainty.\\
    \indent Until recently
    \citep{2009-LPI-Schmidt,2009-Icarus-Drummond}, 
    the volume of Pallas was poorly constrained. The IRAS
    measurement led to a density of
    $\rho_{\textrm{\tiny{IRAS}}} = 3.7 \pm 1.1$ g.cm$^{-3}$,
    and not enough occultations were observed to derive an accurate
    volume \citep[see][]{1989-Icarus-78-Drummond}.
    The density derived here agrees with
    \citet{2009-Icarus-Drummond}
    at the 5\% level, but is about 20\% higher than that determined by 
    \citet{2009-LPI-Schmidt} 
    due to the differences in measured dimensions
    (see above).
    Making further improvements on the density determination will now require
    improved mass estimates.\\
    \indent The difference between the density of (2) Pallas
    ($3.4 \pm$ 0.9 g.cm$^{-3}$) 
    and that of (1) Ceres
    \citep[$\sim$2.2 g.cm$^{-3}$,][]{2008-AA-478-Carry}
    presents a bit of a puzzle.
    Because Ceres and Pallas have been predicted to
    present almost no
    macro-porosity \citep{2002-AsteroidsIII-4.2-Britt}, their bulk
    densities should reflect something close to the mineral density.
    This difference suggests a compositional
    mismatch between these two large bodies,
    even though it has been believed
    for years that they have a similar composition
    \citep[\textsl{e.g.,}][]{1983-Icarus-56-Larson}, close to that
    of carbonaceous chondrites.
    The orbit of Pallas, however,
    being more eccentric 
    than that of Ceres, has a
    perihelion that is
    closer to the Sun by 0.4 AU than the perihelion of Ceres.
    Ceres may thus have retained more
    hydrated (and less dense) materials, as is generally
    proposed to explain its low density
    \citep[\textsl{e.g., see}][]{2005-JGR-110-McCord}. 
    It is also possible that
    Ceres may retain reservoirs of water ice and/or may have a
    somewhat different 
    internal structure than Pallas.
    For example, the near-surface of Ceres may support extensive
    voids relative to Pallas,
    resulting from sublimation of sub-surface
    water ice,
    as predicted by the models of its internal structure
    \citep{1989-Icarus-82-Fanale}.
    Marginal detection of sublimation was claimed by
    \citet{1992-Icarus-98-AHearn},
    although more recent observations 
    \citep{2008-ACM-Rousselot} do not support this idea.
    Based on its near-infrared spectrum, Pallas appears to lack a
    signature 
    of organic or icy material.
    \citep{1990-Icarus-88-Jones}.  We suggest that the sum of
    the evidence points to a dry Pallas, relative to Ceres.

\section{Surface mapping\label{sec-surface}}
  \indent As highlighted in \citet{SurfaceMapping}, the best way to
  study planetary landmarks is to produce surface maps.
  It allows location and comparison of features between independent
  studies and allows correction of possible artifacts (\textsl{e.g.}, from
  deconvolution).
  Here we do not describe the whole
  process of extracting surface maps from AO asteroid images,
  because it has been covered previously for Ceres 
  \citep{2008-AA-478-Carry}. Instead, we report below the main
  improvements with respect to our previous study. 
  \subsection{Method\label{subsec-mapping-method}}
%
%
%
    \paragraph*{Geometry:}~Because a 3-D surface
    cannot be mapped onto a plane
    without introducing distortions, the projection choice
    is crucial, and depends on the geometry of the observations.
    Due to the high obliquity of Pallas (84\degr)
    and its inclined orbit (35\degr), 
    the observations presented here span almost the entire latitude range.
    Following the recommendation of
    \citet{SurfaceMapping}, we produced one
    map for the equatorial band (Equidistant Cylindrical
    Projection) and two others for the polar regions
    (Orthographic Projection),
    thereby minimizing distortion
    over the entire surface of Pallas.
    We used the \citet{2006-arXiv-Goldberg}\footnotemark[3] mapping flexion
    quantification method
    to choose both projections for this specific observation geometry.
\footnotetext[3]{Also available on the web at
  \href{http://www.physics.drexel.edu/~goldberg/projections/}%
       {http://www.physics.drexel.edu/$\sim$goldberg/projections/}}
%
%
%
%
    \paragraph*{Region of interest:}~We decided to exclude
    from the maps the outer annulus of the apparent disk of the asteroid
    in each image.  We did this for two reasons: 1) the image scale
    (km/pixel) and spatial resolution are degraded there
    \citep[][section 4.2]{2008-AA-478-Carry}
    and 2) the edges of
    many of the images suffer from brightness-ringing artifacts resulting from
    deconvolution.
    We defined a 
    Region Of Interest (ROI) to select the range of pixels to be used for 
    mapping. The ROI was defined by the projected shape of Pallas,
    reduced to a given percent of its radius to exclude any artifact. 
    We defined the percentage for each night
    by inspection of the degree of ringing present after image
       restoration. The resulting ROI percentages are given in
     Table~\ref{tab-obs-settings}.
%
%
%
    \paragraph*{Definition of the planetocentric coordinate
      system:}~We define here, for the first time, 
      a planetocentric coordinates system for Pallas,
      following the guidelines of the IAU
      Working Group on cartographic coordinates and rotational elements
      \citep{2007-CeMDA-98-Seidelmann}.
      Longitudes are measured from 0\degr~to 360\degr, following the
      right-hand rule with respect to the spin vector.
      The prime meridian is aligned with
      the long axis, 
      pointing toward negative $x$ in the 
      shape model reference frame.
      Latitudes are measured $\pm$90\degr~from
      the equator,
      with +90\degr~being in the direction of the spin vector.

%
%
    \paragraph*{Projection:}~We used the shape model of Pallas (see
    section \ref{subsec-pallas_model}) to
    convert image pixels to their prints on
    the surface of Pallas.
    For each image, we produced an equivalent image of the
    shape model projected onto the plane of the sky. 
    We then derived the planetocentric coordinates of each
    pixel (longitude and latitude). Finally, we convert those
    planetocentric coordinates to x, y positions on the map,
    using the translation equations appropriate for the
    particular map projection.
%
%
%
%
    \paragraph*{Combination of images into maps:}~There was no overlap
    between the northern and southern hemispheres
    in our data. Therefore, we had to
    arbitrarily set their relative brightness to produce a complete
    map. Ultimately, we assumed
    both hemispheres to have the same mean albedo, because no
    evidence for such a difference exists in the literature.
    To handle redundant coverage, \textsl{i.e.}, where more than one
    image covers a specific region, 
    we use an average of all the images, with higher weight
    given to higher
    resolution
    and/or higher quality images
    \citep[see][for detailed explanation]{2008-AA-478-Carry}.
  \subsection{(2) Pallas surface in the near-infrared\label{subsec-mapping-description}}
    \indent Because the J and H filters
    were used only sparsely (Table~\ref{tab-obs-condition}),
    the K-band map covers a larger fraction
     of the Pallas surface
    (80\% for K \textsl{vs.} 40\% for J and H).
    So limited imaging can, of course, restrict the explored
    area on the asteroid surface; but also, fewer
    overlapping 
    images of one area will result in greater errors than
    for regions that have a larger number of redundant images.

    As explained in section~\ref{sssec-contour},
    the deconvolution process
    can lead to the creation of artifacts. 
    Although we rejected deconvolved images of poor quality and
    re-applied the \textsc{Mistral} 
    deconvolution process until the dataset was self-consistent, the final 
    products can still show discrepancies
     between images of the same 
    region of Pallas 
    (e.g., introduced by the incomplete AO correction).
    The best way to smooth out such artifacts is to
    combine as many images as possible, and use their
    mean value to produce the final maps. 
    This method assumes that the probability of recovering real
    information is greater than the 
    probability of introducing additional artifacts 
    with \textsc{Mistral}.
    This assumption is increasingly valid with increasing 
    signal-to-noise ratio and increasing number of overlapping images
    (our observations are optimized to provide
    high signal-to-noise, usually at levels
    of several hundred). 

    \indent An additional test of the validity of the
    \textsc{Mistral} deconvolution comes from the comparison of
    AO-VLT deconvolved images of bodies also observed \textsl{in
      situ} by spacecraft.
    Ground-based observations of Jupiter's moon Io
    \citep{2002-Icarus-160-Marchis} and Saturn's moon Titan
    \citep{2006-JGR-111-Witasse} have been found to be in good
    agreement with sizes measured from Galileo and Cassini spacecraft
    data.\\ 
%
%
%
%
%
\begin{figure*}
\begin{center}

  \textbf{K-band maps}\\
  \includegraphics[width=.4\textwidth]{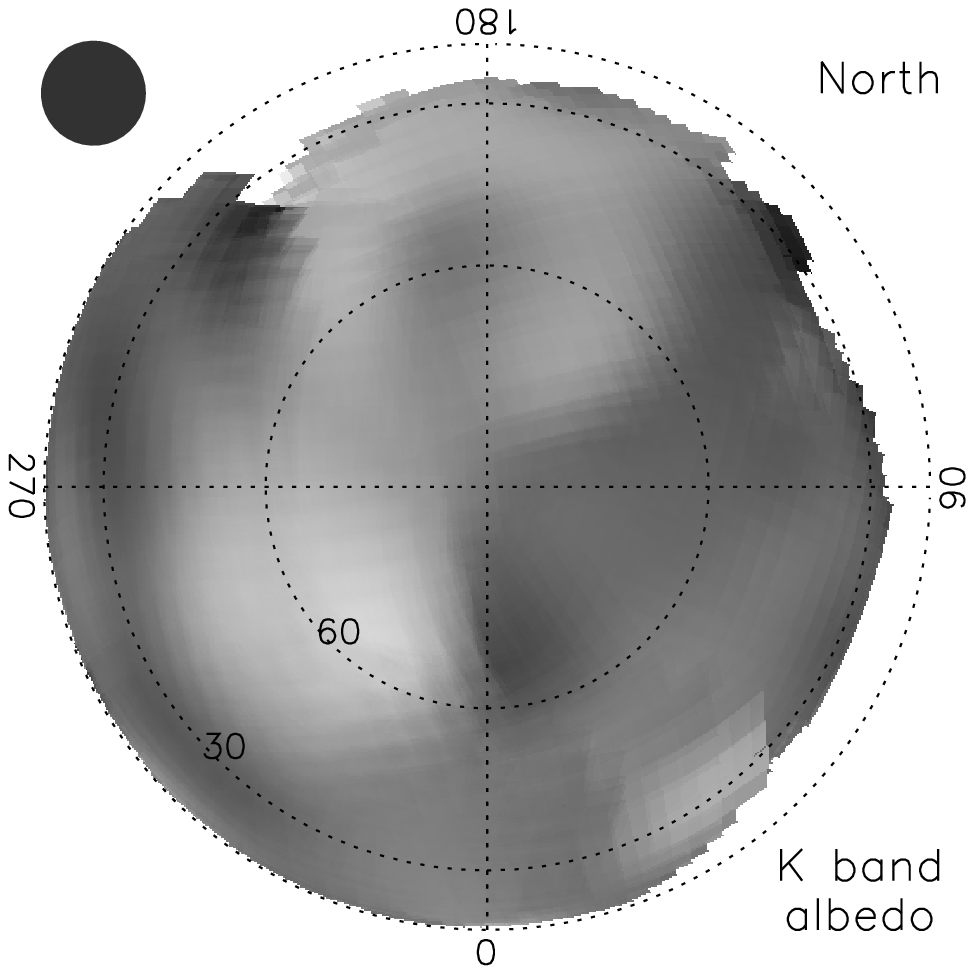}%
  \includegraphics[width=.4\textwidth]{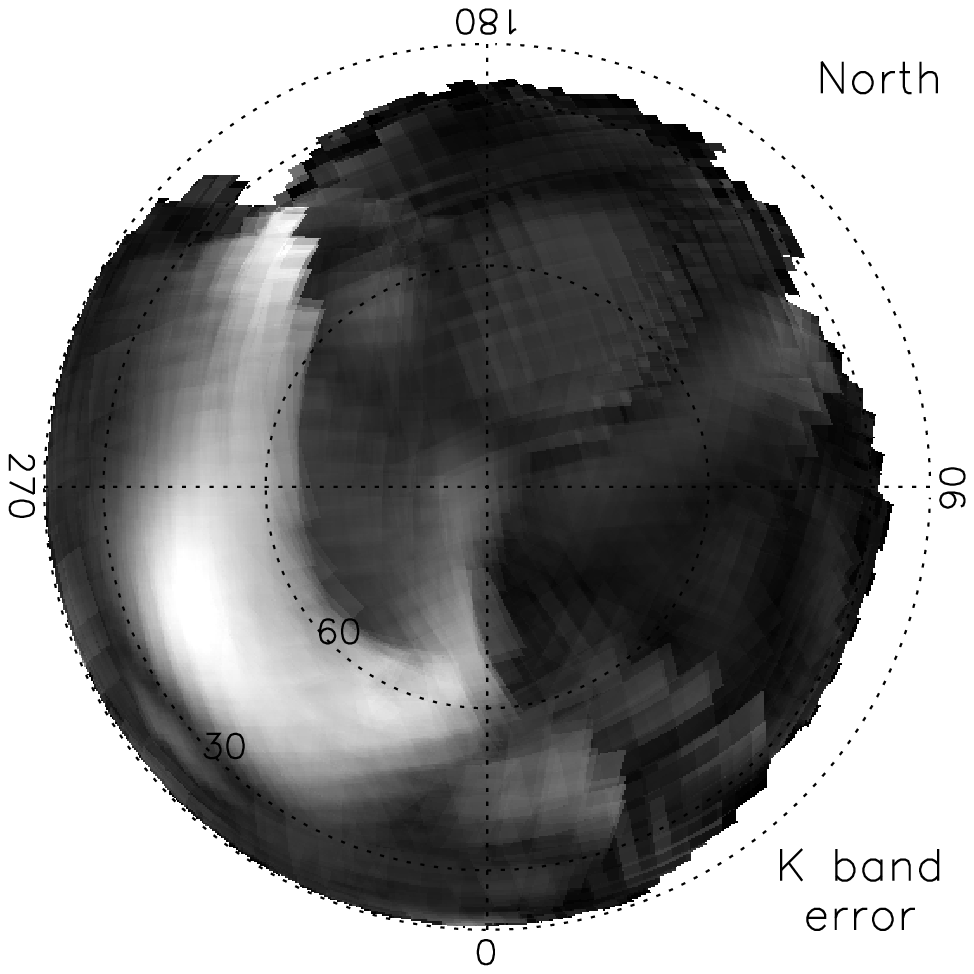}\\
  \includegraphics[width=.8\textwidth]{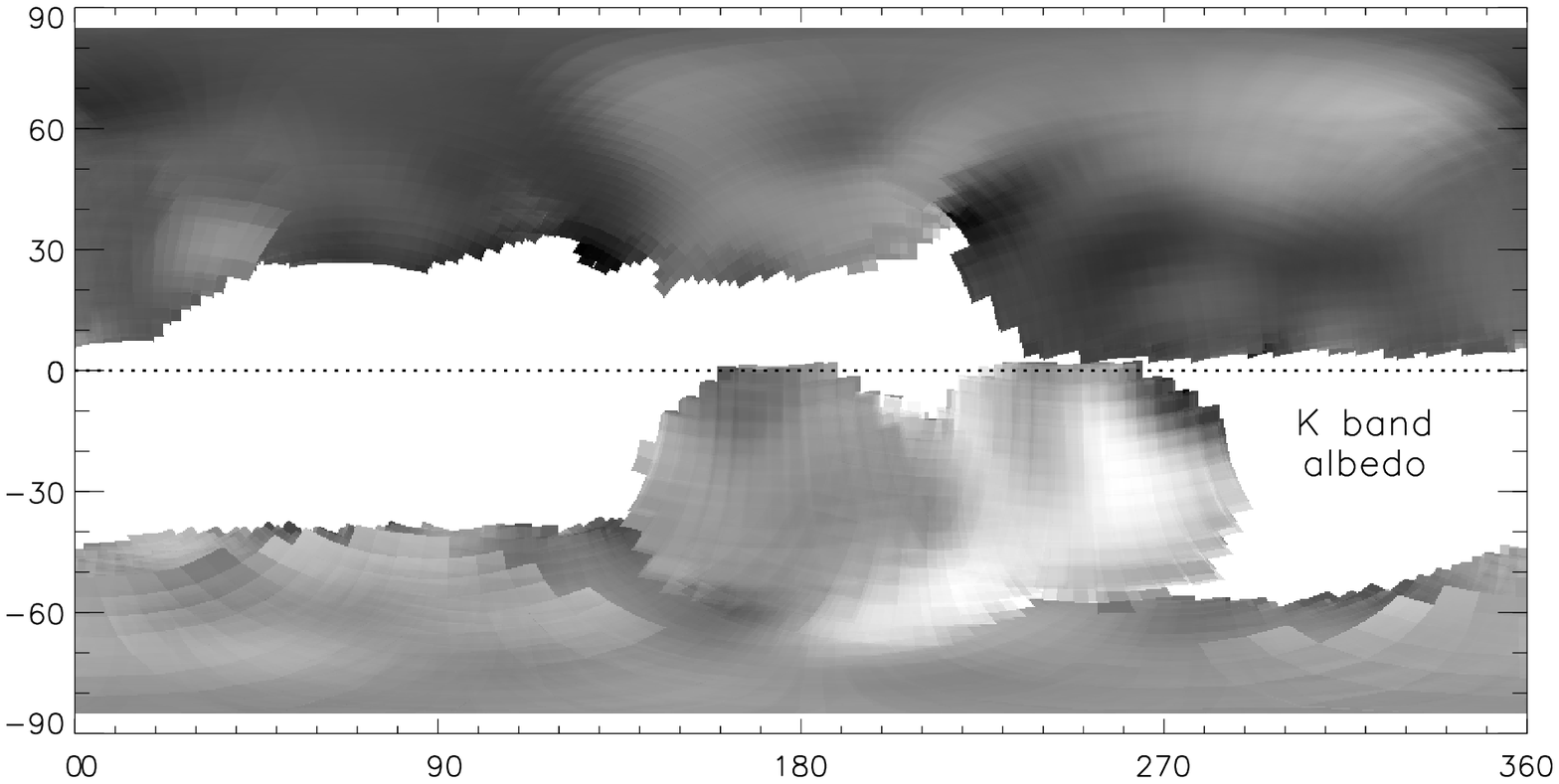}\\
  \includegraphics[width=.4\textwidth]{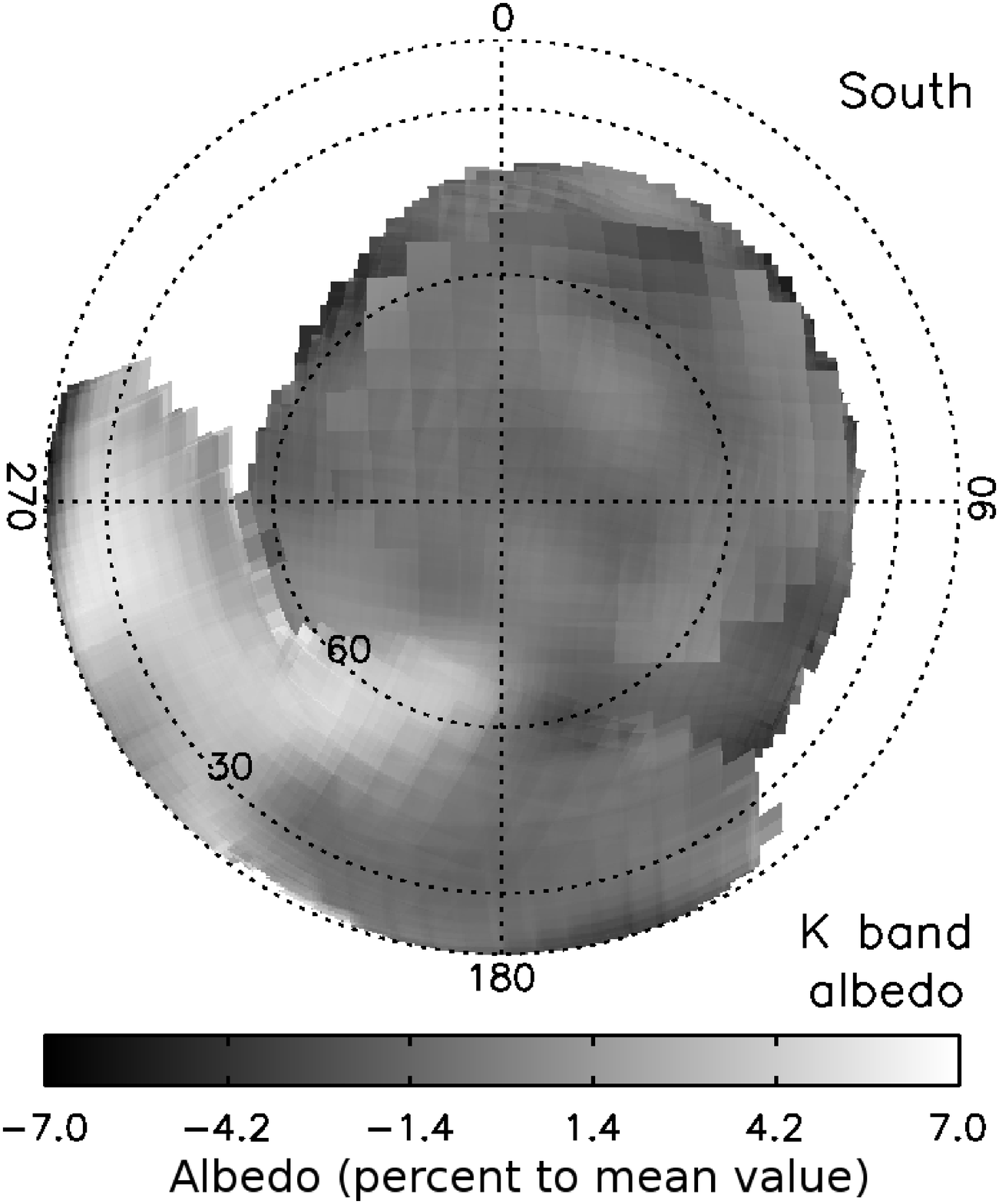}%
  \includegraphics[width=.4\textwidth]{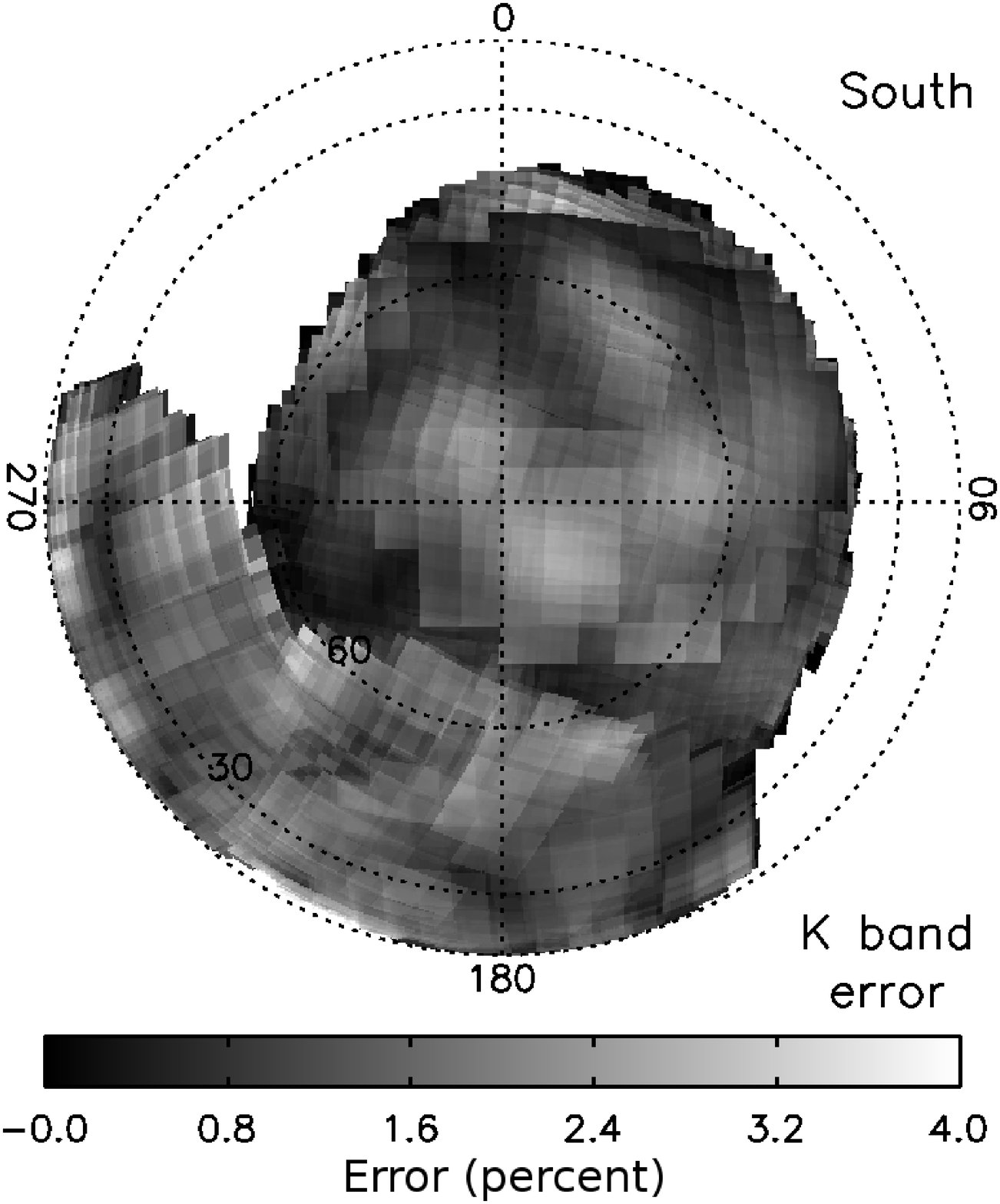}\\

  \caption[\textsl{K}-band Map]{
    \textsl{K}-band map covering $\sim$80\% of Pallas' surface. The areas in
    white are \textsl{terra incognita} due to the limited ROI (see
    text) and
    Pallas' spatial orientation during the observations
    (Table.~\ref{tab-obs-condition}), including most of the
    equatorial regions.
    The final estimated resolution element $\Theta$ for this
    composite map is
    shown at the upper-left corner.
    The albedo varies by about
    $\pm$6\% around the mean surface value for each map. We estimate
    the errors to be limited to 4\% maximum.
    The grey scale is common for
    the three maps.
    The mapping for the southern hemisphere in general appears coarser than
    for the northern hemisphere, due to the
    smaller number of
    available images for the southern hemisphere.
    Any feature or albedo distribution present in
    these maps has a 
    very low probability of being an artifact.
    \label{fig-map.K}
    \label{lastfig}}

  \end{center}
\end{figure*}

\begin{figure*}
\begin{center}

  \textbf{J- and H-band maps}\\
  \includegraphics[width=.5\textwidth]{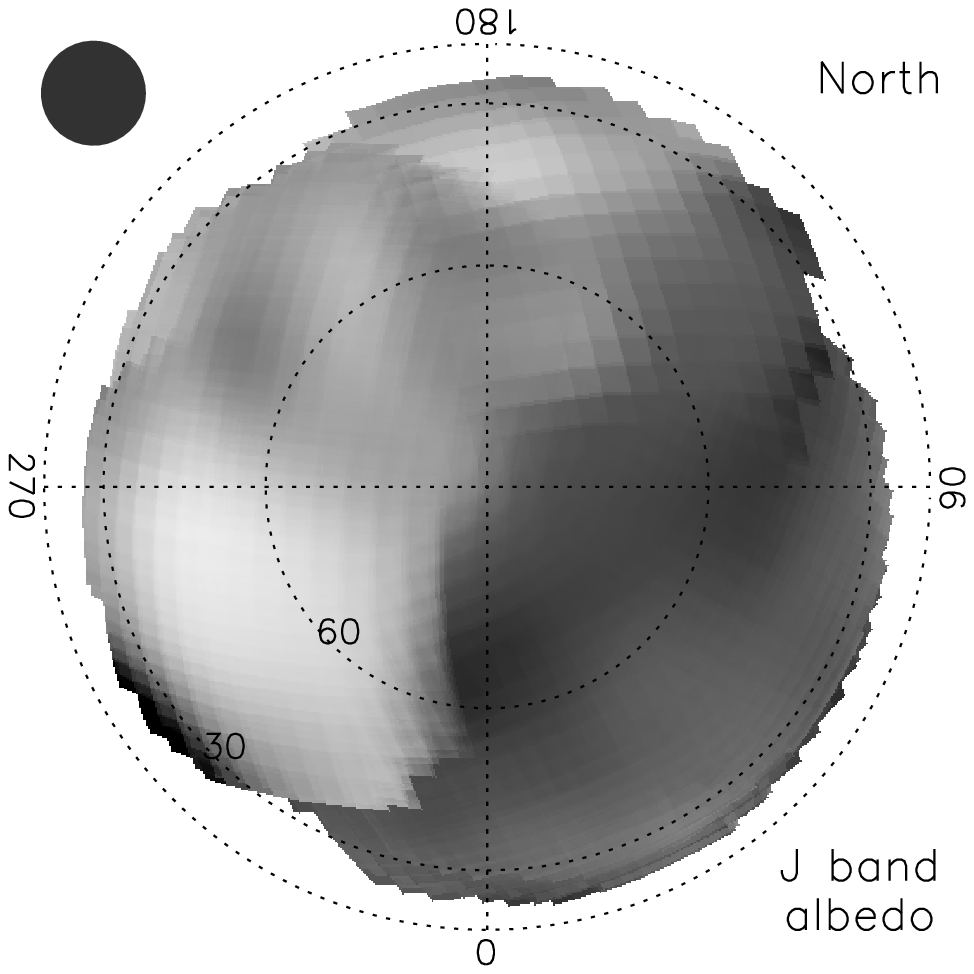}%
  \includegraphics[width=.5\textwidth]{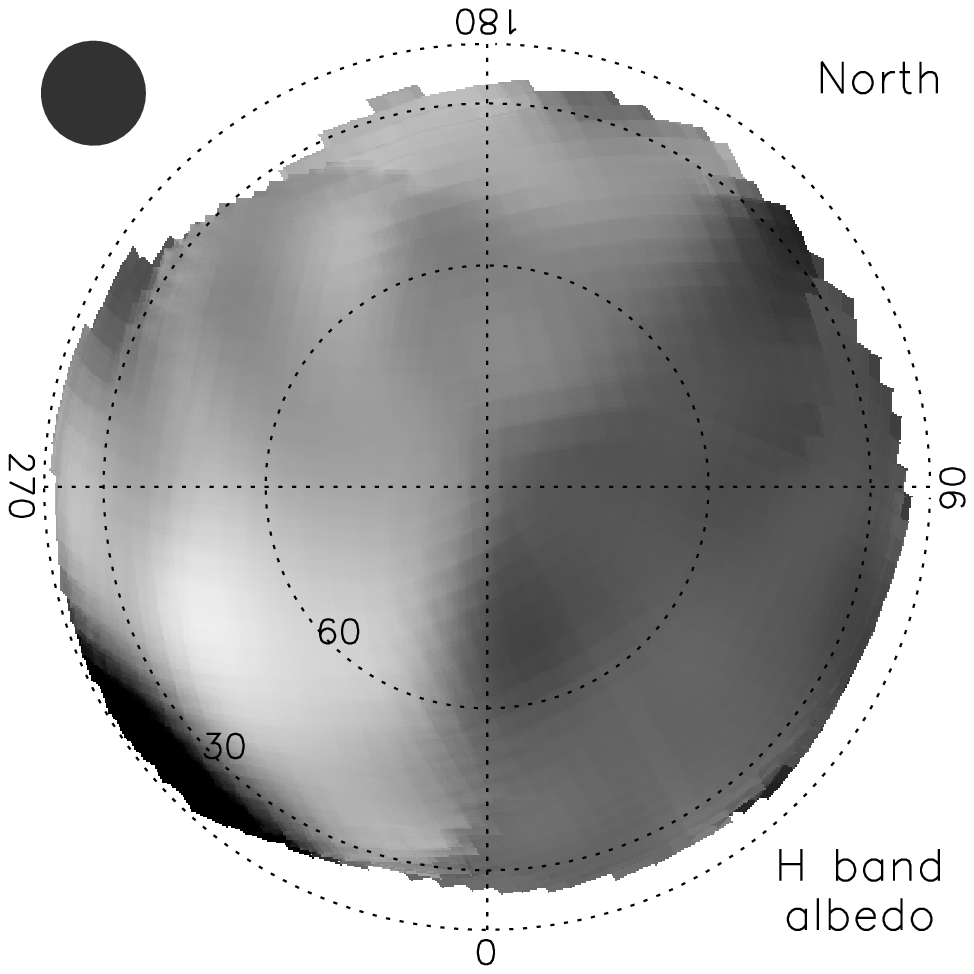}\\
  \includegraphics[width=.5\textwidth]{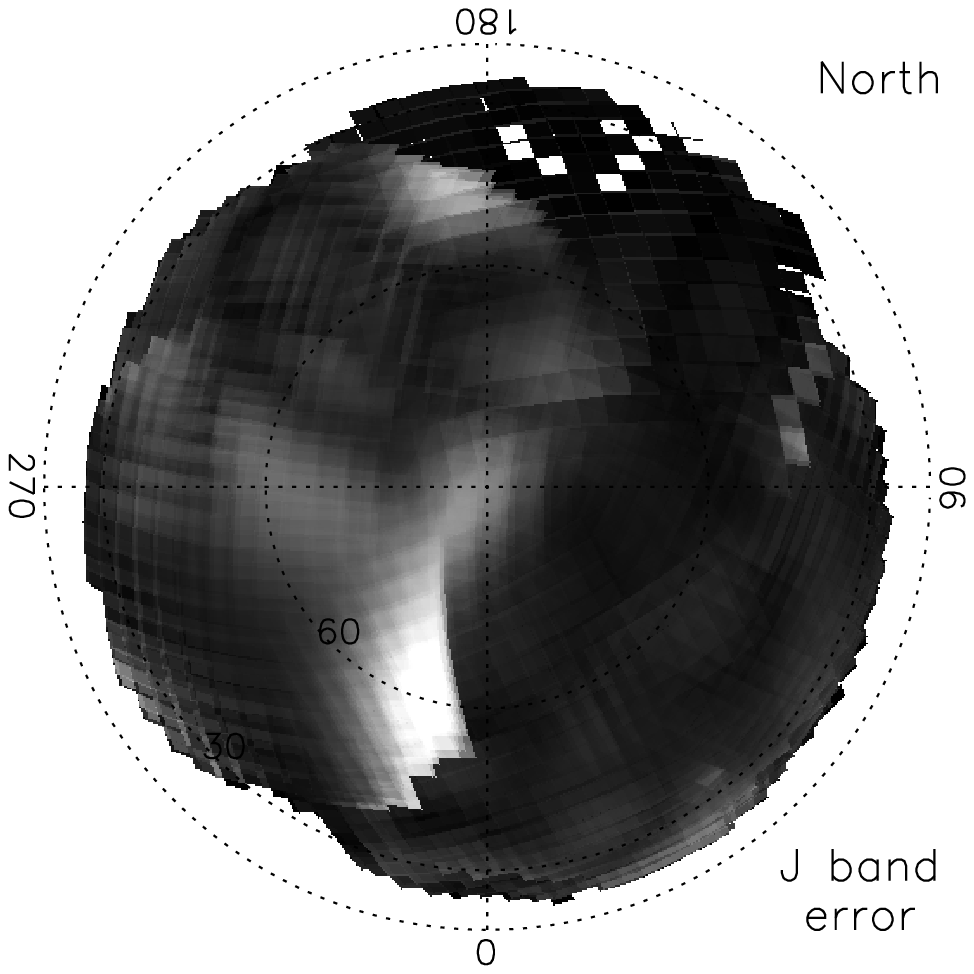}%
  \includegraphics[width=.5\textwidth]{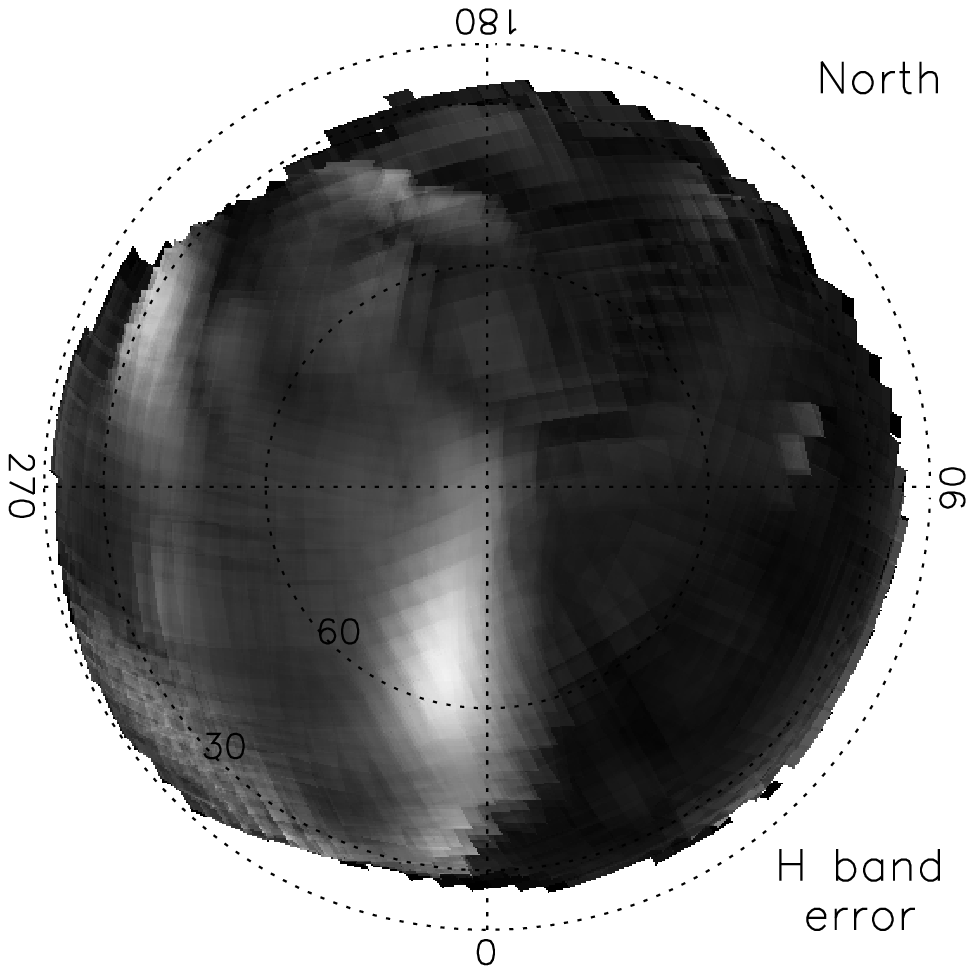}\\


  \caption[J- and H-band Map]{
    Same as for the K-band observations (Fig.~\ref{fig-map.K}).
    The coarser appearance of these maps results
    from fewer images being used to produce the maps
    (27 for J and 44 for H, \textsl{vs}. 115 for K).
    We only produced an Orthographic projection for
    these filters because
    the southern hemisphere was not imaged in these bands.
    As for the K-band, the albedo variations are within
    $\pm$6\% of the mean surface value for each map and we estimate
    the errors to be limited to 4\% maximum. The final estimated resolution
    elements for the composites are shown also to the upper left of each map, as
    in Fig.~\ref{fig-map.K}.
    \label{fig-map.JH}
    \label{lastfig}}

  \end{center}
\end{figure*}
%
%
%
    \indent The \textsl{J}-, \textsl{H}- and \textsl{K}-band maps shown in
    Fig.~\ref{fig-map.K} and Fig.~\ref{fig-map.JH} are the
    result of combining 27, 44 and 115 individual projections,
    respectively.
    The spatial resolution for these composite maps is nearly
    equivalent across the three bands, and is $\sim$60 km.
    The amplitude of the albedo variation is within $\pm$6\% of the
    mean surface value for each band. 
    From the albedo error maps (obtained by measuring, for each
    pixel, the intensity dispersion among the individual maps), we
    report a maximal error of $\sim$4\% (mean error is below 2.5\%).\\
    \indent Pallas shows a large, dark region between
    0\ensuremath{^\circ}~and $\sim$120\ensuremath{^\circ}~in longitude
    in the northern hemisphere, 
    where the shape model presents a facet or ``depression''.
    The fact that we see this feature
    at all wavelengths suggests that it is real and could be associated
    with a geological feature such  
    as an impact crater.
    However, because the light-curve inversion was done without taking
    into account the
    albedo information, the depression seen in our model may be an
    artifact created by the light-curve inversion algorithm 
    (as suggested by the occultation chords, see Fig.~\ref{fig-occ}).
    In future versions of the KOALA method, we will attempt to 
    use the albedo information from the images to improve the shape
    model.\\
    \indent Some other features are remarkable, such as the dark spot
    (diameter $\sim$70 km)
    surrounded by a bright annulus 
    (about 180 km at its largest extent) at
    (185\ensuremath{^\circ}, +50\ensuremath{^\circ}) or the 
    bright region (diameter of $\sim$110 km) around
    (300\ensuremath{^\circ}, +60\ensuremath{^\circ}). 
    Southern features are more difficult to interpret because of the
    higher noise and the lack of observations in J- and H-band that
    preclude a cross-check with the
    features in the K-band observations.\\
    \indent The surface of Pallas
    appears to have fewer small-scale
    structures (of size comparable to the resolution element)
    than Ceres 
    \citep{2006-Icarus-182-Li, 2008-AA-478-Carry}, even though
    both objects were observed at approximately the same spatial
    resolution.
    Similarly, 
    Vesta also does not exhibit
    small-scale features when observed at comparable spatial
    resolution
    \citep[see][]{1997-Icarus-128-Binzel, 2008-ACM-Li}.\\
    \indent To look for color variations, we also selected several
    regions in the northern hemisphere
    (three dark and four bright)
    and measured their relative flux in the three wavebands.
    As a result, we detect spectral variations slightly above
    the noise level, but without remarkable behavior.
    These differences could be due to morphological features or
    differences in the
    surface composition and/or regolith properties (such as grain
    size).
    One could interpret these variations as minealogical heterogeneity, but
    the differences are weak with the existing dataset.

\section{Conclusion}
  \indent We report here the first study of an asteroid using a
  new approach combining light-curves and occultation data with 
  high-angular resolution
  images obtained with adaptive optics (AO), which we have termed
  KOALA for Knitted
  Occultation, Adaptive optics and Light-curve Analysis.
  This method allows us to derive the spin vector coordinates,
  and to produce an \textsl{absolute-sized} shape model of the
  asteroid, providing an improved volume measurement.
  This method can be used on any body for which light-curves and
  disk-resolved images are
  available at several geometries.

  \indent Here, we analyze all the 
  near-infrared high-angular resolution images of
  Pallas that we
  acquired from 2003 to 2007.
  We find the
  spin vector coordinates of Pallas to be
  within 5\ensuremath{^\circ}~of 
  (30\ensuremath{^\circ}, -16\ensuremath{^\circ}) in the
  Ecliptic J2000.0
  reference frame, indicating
  a high obliquity of about 84\ensuremath{^\circ} and implying
  large seasonal effects on Pallas.\\
  \indent The derived shape model 
  reproduces well both the Pallas' projected outline
  on the sky and its light-curve
  behavior at all epochs.
  Our best-fit tri-axial ellipsoid radii are
  $a$=275 $\pm$ 4 km, $b$= 258 $\pm$ 3 km, and $c$= 238 $\pm$ 3 km,
  allowing us to estimate an average density for Pallas
  of 3.4 $\pm$ 0.9 g.cm$^{-3}$
  \citep[using M=(1.2 $\pm$ 0.3) $\times$ 10$^{-10}$
      M$_\odot$ from][]{2000-AA-360-Michalak}.
  The density uncertainty is now almost entirely due to mass uncertainty.
  This density might be interpreted as a result of a dryer Pallas with
  respect to Ceres (supported by spectroscopic studies).

  The observation of such a large difference in the bulk
  density of two large asteroids of similar taxonomic type, of 
  apparently similar surface compositions, and apparently
  lacking in significant macro-porosity, underscores the need for
  dedicated programs to monitor
  close 
  encounters between asteroids
  \citep[\textsl{e.g.,} from GAIA
    observations,][]{2007-AA-472-Mouret}, in 
  turn allowing us to
  derive more accurate masses and improve our knowledge of 
  asteroid densities.\\
  \indent We also present the first albedo maps of Pallas,
  revealing  features with diameters in the 70$-$180 km range and an albedo
  contrast of about 6\% with respect to the mean surface
  albedo.
  Weak spectral variations are also reported.

\section*{Acknowledgements}
We would like to thank Franck Marchis (SETI Institute) for the flat-field frames he
provided for our August 2006 observations.
Thanks to Team Keck for their support and Keck Director Dr. Armandroff
for the use of NIRC2 data obtained on 2007 July 12 technical time.
Partial support for this work was provided by NASA's Planetary
Astronomy Program (PIs Dumas and Merline), NASA's OPR Program (PI Merline)
and NSF's Planetary Astronomy Program (PI Merline).
M.K. was supported by
the Academy of Finland (project: New mathematical methods in planetary
and galactic research).
Thanks to Bill Bottke (SwRI), Anne Lema\^{i}tre (University
Notre-Dame de la Paix) and Ricardo Gil-Hutton (San Juan
University) for discussions on Pallas.
Thanks also to Francesca Demeo (Observatoire de Paris) for her careful
reading of this article 
and the correction to the English grammar.
Thanks to both anonymous referees who provided constructive
comments on this article.
The authors wish to recognize and acknowledge the very significant
cultural role and reverence that the summit of Mauna Kea has always
had within the indigenous Hawaiian community.  We are most fortunate
to have the opportunity to conduct observations from this mountain.

\label{lastpage}





\end{document}